\documentclass[twocolumn]{aastex62}
\usepackage[figuresright]{rotating}

\submitjournal{ApJS}

\begin{document}
\title{Tracing Kinematic and Chemical Properties of Sagittarius Stream by K-Giants, M-Giants, and BHB stars}

\correspondingauthor{Xiang-Xiang Xue; Chengqun Yang} \email{xuexx@nao.cas.cn; ycq@bao.ac.cn}

\author[0000-0003-1972-0086]{Chengqun Yang}
\affiliation{National Astronomical Observatories, Chinese Academy of Sciences, 20A Datun Road, Chaoyang District, Beijing 100101, P.R.China}
\affiliation{School of Astronomy and Space Science, University of Chinese Academy of Sciences, 19A Yuquan Road, Shijingshan District, Beijing 100049, P.R.China}

\author[0000-0002-0642-5689]{Xiang-Xiang Xue}
\affiliation{National Astronomical Observatories, Chinese Academy of Sciences, 20A Datun Road, Chaoyang District, Beijing 100101, P.R.China}

\author[0000-0002-4953-1545]{Jing Li}
\affiliation{Physics and Space Science College, China West Normal University, 1 ShiDa Road, Nanchong 637002, P.R.China}
\affiliation{Chinese Academy of Sciences South America Center for Astronomy, National Astronomical Observatories, CAS, Beijing 100012, China}

\author[0000-0002-1802-6917]{Chao Liu}
\affiliation{National Astronomical Observatories, Chinese Academy of Sciences, 20A Datun Road, Chaoyang District, Beijing 100101, P.R.China}
\affiliation{School of Astronomy and Space Science, University of Chinese Academy of Sciences, 19A Yuquan Road, Shijingshan District, Beijing 100049, P.R.China}

\author[0000-0002-6434-7201]{Bo Zhang}
\affiliation{National Astronomical Observatories, Chinese Academy of Sciences, 20A Datun Road, Chaoyang District, Beijing 100101, P.R.China}
\affiliation{School of Astronomy and Space Science, University of Chinese Academy of Sciences, 19A Yuquan Road, Shijingshan District, Beijing 100049, P.R.China}

\author[0000-0003-4996-9069]{Hans-Walter Rix}
\affiliation{Max-Planck-Institute for Astronomy K\"{o}nigstuhl 17, D-69117, Heidelberg, Germany}

\author{Lan Zhang}
\affiliation{National Astronomical Observatories, Chinese Academy of Sciences, 20A Datun Road, Chaoyang District, Beijing 100101, P.R.China}

\author[0000-0002-8980-945X]{Gang Zhao}
\affiliation{National Astronomical Observatories, Chinese Academy of Sciences, 20A Datun Road, Chaoyang District, Beijing 100101, P.R.China}
\affiliation{School of Astronomy and Space Science, University of Chinese Academy of Sciences, 19A Yuquan Road, Shijingshan District, Beijing 100049, P.R.China}

\author[0000-0003-3347-7596]{Hao Tian}
\affiliation{National Astronomical Observatories, Chinese Academy of Sciences, 20A Datun Road, Chaoyang District, Beijing 100101, P.R.China}

\author[0000-0001-5245-0335]{Jing Zhong}
\affiliation{Key Laboratory for Research in Galaxies and Cosmology, Shanghai Astronomical Observatory, Chinese Academy of Sciences, 80 Nandan Road, Shanghai 200030, People’s Republic of China}

\author[0000-0003-0663-3100]{Qianfan Xing}
\affiliation{National Astronomical Observatories, Chinese Academy of Sciences, 20A Datun Road, Chaoyang District, Beijing 100101, P.R.China}

\author{Yaqian Wu}
\affiliation{National Astronomical Observatories, Chinese Academy of Sciences, 20A Datun Road, Chaoyang District, Beijing 100101, P.R.China}

\author{Chengdong Li}
\affiliation{Rudolf Peierls Centre for Theoretical Physics, Clarendon Laboratory, Parks Road, Oxford OX1 3PU, UK}

\author[0000-0002-3936-9628]{Jeffrey L. Carlin}
\affiliation{LSST, 950 North Cherry Avenue, Tucson, AZ 85719, USA}

\author{Jiang Chang}
\affiliation{National Astronomical Observatories, Chinese Academy of Sciences, 20A Datun Road, Chaoyang District, Beijing 100101, P.R.China}


\begin{abstract}
We characterize the kinematic and chemical properties of $\sim$3,000 Sagittarius (Sgr) stream stars, including K-giants, M-giants, and BHBs, select from SEGUE-2, LAMOST, and SDSS separately in Integrals-of-Motion space. The orbit of Sgr stream is quite clear from the velocity vector in $X$-$Z$ plane. Stars traced by K-giants and M-giants present the apogalacticon of trailing steam is $\sim$ 100 kpc.
The metallicity distributions of Sgr K-, M-giants, and BHBs present that the M-giants are on average the most metal-rich population, followed by K-giants and BHBs. All of the K-, M-giants, and BHBs indicate that the trailing arm is on average more metal-rich than leading arm, and the K-giants show that the Sgr debris is the most metal-poor part.
The $\alpha$-abundance of Sgr stars exhibits a similar trend with the Galactic halo stars at lower metallicity ([Fe/H] $<\sim$ $-$1.0 dex), and then evolve down to lower [$\alpha$/Fe] than disk stars at higher metallicity, which is close to the evolution pattern of $\alpha$-element of Milky Way dwarf galaxies.
We find $V_Y$ and metallicity of K-giants have gradients along the direction of line-of-sight from the Galactic center in $X$-$Z$ plane, and the K-giants show that $V_Y$ increases with metallicity at [Fe/H] $>\sim-$1.5 dex.
After dividing the Sgr stream into bright and faint stream according to their locations in equatorial coordinate, the K-giants and BHBs show that the bright and faint stream present different $V_Y$ and metallicities, the bright stream is on average higher in $V_Y$ and metallicity than the faint stream.
\end{abstract}



\defcitealias{LM10}{LM10}
\defcitealias{DL17}{DL17}

\section{Introduction}\label{sec:intro}
The disrupting Sagittarius (Sgr) dwarf spheroidal galaxy (dSph) was discovered in the work on the Galactic bulge of \citet{Ibata94}, which has a heliocentric-distance of $\sim$25 kpc and is centered at coordinate of $l=5.6^\circ$ and $b=-14.0^\circ$ \citep{Ibata97}. For a dwarf galaxy, such a close distance to the Galactic center means it is suffering a huge tidal force from the Milky Way. Subsequently, the Sgr stream was found \citep{Yanny00, Ibata01} and has been traced over 360$^\circ$ on the Sky \citep{Majewski03, Belokurov06}, which indicates it is a strong tool for exploring the Milky Way \citep{Ibata97, Majewski99}.

Thanks to the early detections of Sgr tidal stream from Two Micron All Sky Survey \citep[2MASS;][]{Skrutskie06} and Sloan Digital Sky Survey \citep[SDSS;][]{York00}, the morphology of the Sgr stream is known in detail \citep{Newberg02, Newberg03, Majewski03, Majewski04}. Earlier model and observations predicted the Galactocentric distance of the Sgr stream about $\sim$20-60 kpc \citep{LM10, Majewski03}. Recently, \citet{Belokurov14}, \citet{Koposov15} and \citet{Hernitschek17} found the Sgr trailing stream reaches $\sim100$ kpc from the Sun. \citet{Sesarl17} and \citet{Li19} used RR Lyrae stars and M-giants found the trailing stream even extends to a heliocentric distance of $\sim$ 130 kpc at $\widetilde{\Lambda}_\odot \sim170^\circ$\footnote{ ($\widetilde{\Lambda} _\odot, \widetilde{B}_\odot)$ used in this paper is a heliocentric coordinate system defined by \citet{Belokurov14}. The longitude $\widetilde{\Lambda}_\odot$ begins at the Sgr core and increases in the direction of the Sgr motion. The equator (latitude $\widetilde{B}_\odot$ = 0$^\circ$) is aligned with the Sgr trailing stream.}. Additionally, \citet{Belokurov06} and \citet{Koposov12} found that the Sgr stream has a faint bifurcation called faint stream, which is a always on the same side of the bright stream at a nearly constant angular separation and without cross \citep{NC16}.

It has been recognized that the Sgr dSph has a complex star formation history. \citet{Ibata95} showed that Sgr contains a strong intermediate-age population with age $\sim4$ - 8 Gyr and metallicity $\sim-$0.2 to $-0.6$ dex and its own globular cluster system. \citet{Siegel07} demonstrated that Sgr has at least 4 - 5 star formation bursts, including an old population: 13 Gyr and [Fe/H] = $-1.8$ dex from main sequence (MS) and red-giant branch (RGB) stars; at least two intermediate-aged populaitons: 4 - 6 Gyr with [Fe/H]= $-0.4$ to $-0.6$ dex from RGB stars; a 2.3 Gyr population near solar abundance: [Fe/H] = $-0.1$ dex from main sequence turn-off (MSTO) stars. \citet{Carlin18} picked up 42 Sgr stream stars from LAMOST M-giants and processed high-resolution observations, they found stars in trailing and leading streams show systematic differences in [Fe/H], and the $\alpha$-abundance patterns of Sgr stream is similar to those observed in Sgr core and other dwarf galaxies like the large Magellanic Cloud and the Fornax dwarf spheroidal galaxy.

With the second data release of $Gaia$ mission \citep{Gaia18}, it is becoming possible to search the Galactic halo substructures in 6D phase space. Xue, X.-X et al. (2019, in preparation, X19 thereafter) took advantage of 6D information to obtain about 3,000 Sgr stream members with high reliability in Integrals-of-Motion (IoM) space, which is the largest spectroscopic Sgr stream sample obtained yet. Based on this sample, we will characterize the properties of the Sgr stream in more detail. This paper is structured as follow: in Section \ref{sec:data&method}, we describe our Sgr sample and the method of X19 used for selecting the Sgr members. In Section \ref{sec:Sgr}, we present the kinematic and chemical properties of the Sgr sample. Finally, a brief summary is shown in Section \ref{sec:summary}.

\section{DATA and METHOD}\label{sec:data&method}
\subsection{Data}
The Sgr stream sample consists of K-, M-giants, and blue horizontal branch stars (BHBs). The K-giants are from Sloan Extension for Galactic Understanding and Exploration 2 \citep[SEGUE-2;][]{Yanny09} and the fifth data release of Large Sky Area Multi-Object Fibre Spectroscopic Telescope \citep[LAMOST DR5;][]{Zhao12, Cui12, Luo12}, and their distances were estimated by Bayesian method \citet{Xue14}. The M-giants are picked up from LAMOST DR5 through a 2MASS+WISE photometric selection criteria. The distances were calculated through the $(J-K)_0$ color-distance relation \citet{Li16, Li19, Zhong19}. The BHBs are chosen from SDSS by color and Balmer line cuts, and their distances were easy to estimate because of the nearly constant absolute magnitude of BHB stars \citep{Xue11}.

\begin{figure}[htb]
\centering
  \begin{tabular}{@{\hspace{-0.5cm}}cccc@{}}
    \includegraphics[width=.50\textwidth]{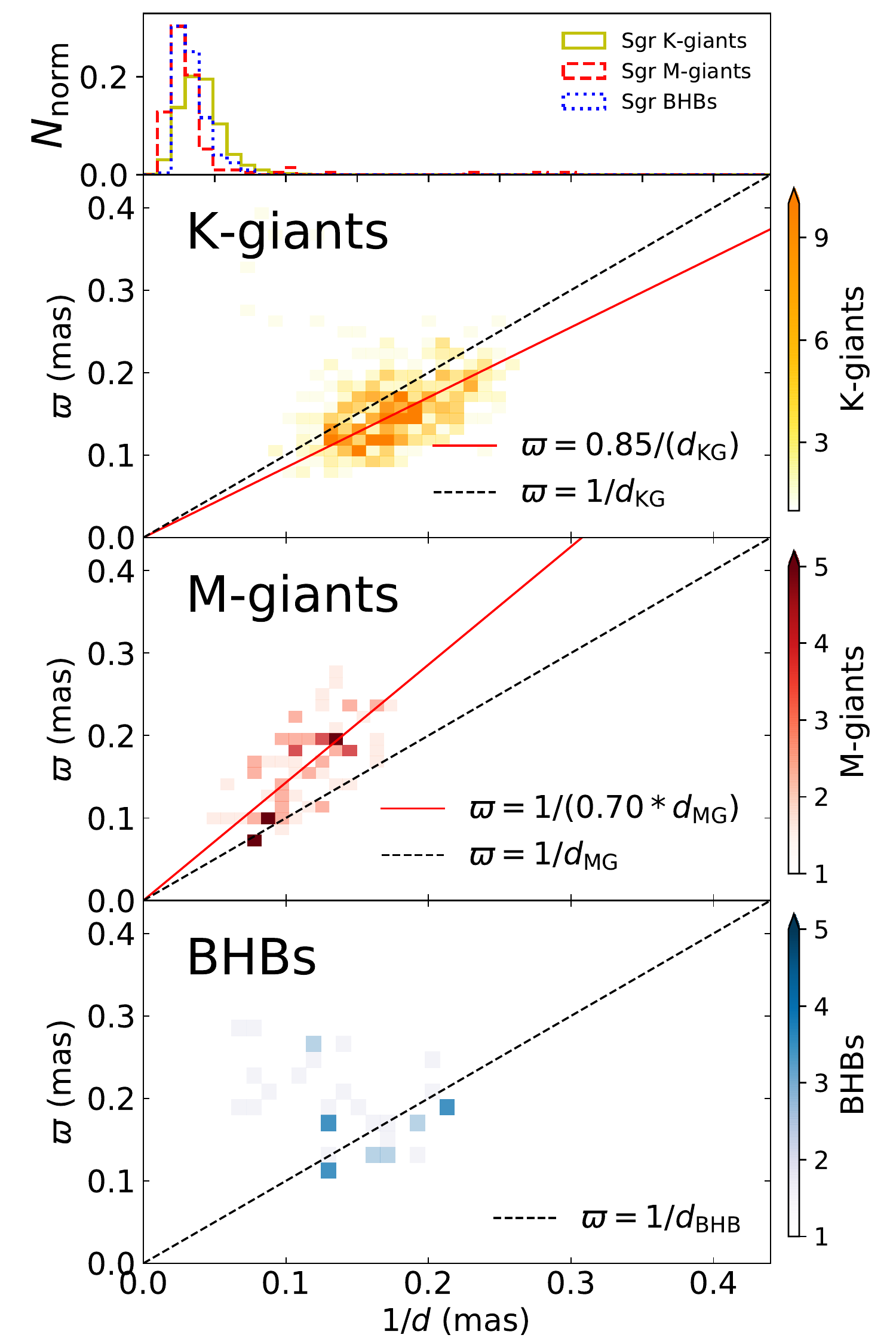} &
    \includegraphics[width=.50\textwidth]{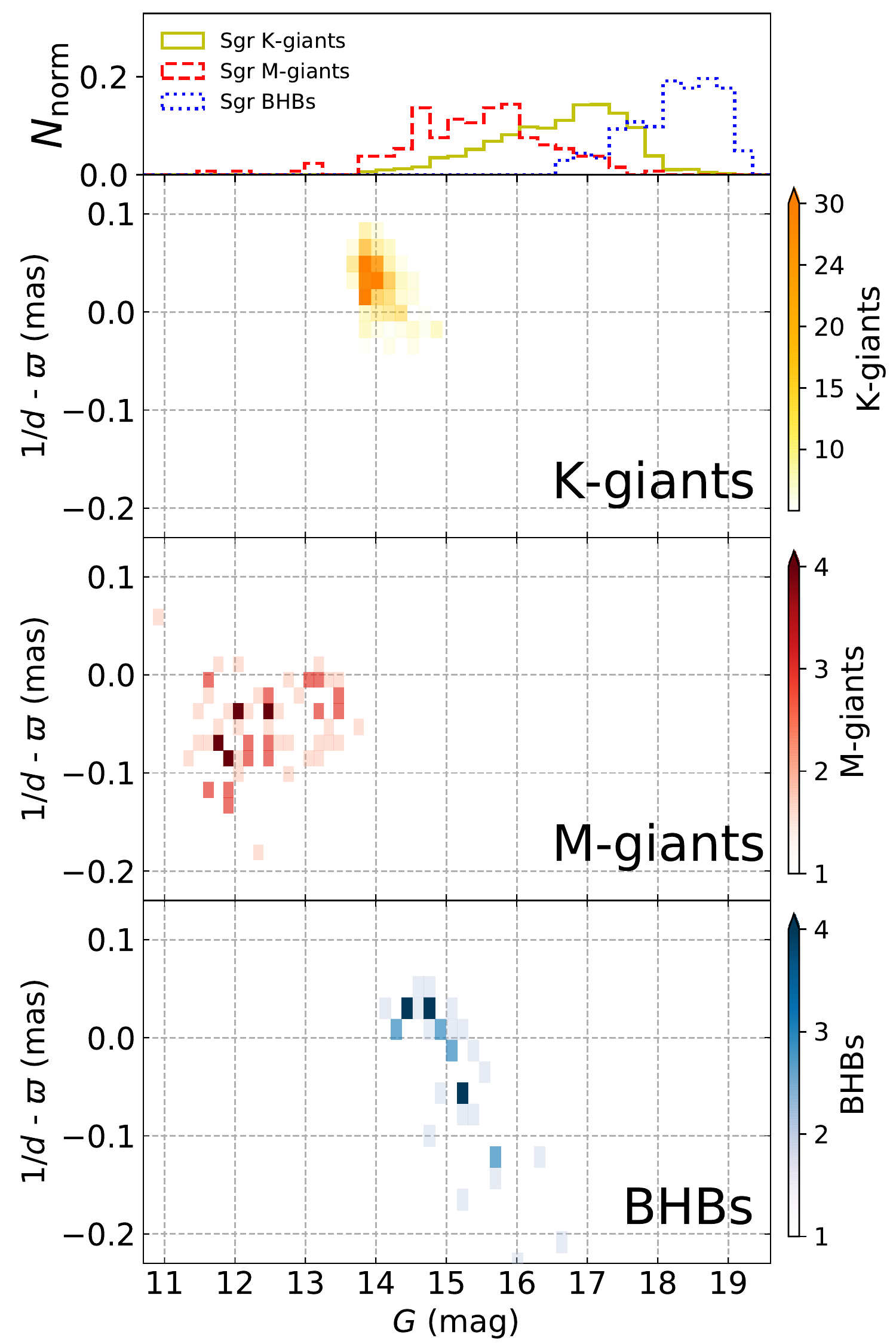} &
    \\
  \end{tabular}
\caption{Top two panels: the inverse of distances $1/d$ and $Gaia\ G$ magnitude distributions of Sgr stars. Histograms with different colors represent different kinds of stars. Left lower three panels: density map comparisons of $Gaia$ DR2 parallax ($\varpi$) with the inverse of distances ($1/d$) of K-giants, M-giants, and BHBs. The black dashed lines mark the 1:1 relation between the scales, and the red lines represent the systematic bias between them, which are fitted by least squares method. Right lower three panels: density map comparisons of parallax biases ($1/d-\varpi$) with $G$ magnitude of K-, M-giants, and BHBs.} \label{sys}
\end{figure}

We calibrated the distances of K-, M-giants, and BHBs with $Gaia$ DR2 parallax rather than Gaia distances estimated by \citet{BJ18}. Because \citet{BJ18} claimed that their mean distances to distant giants are underestimated, because the stars have very large fractional parallax uncertainties, so their estimates are prior-dominated, and the prior was dominated by the nearer dwarfs in the model. Only stars with good parallaxes ($\delta \varpi/\varpi <20\%$) and good distances ($\delta d/d <20\%$) are used to do the calibration, which allows us to compare parallax with $1/d$, and minimize the possible bias from inverting. It is very hard to use Sgr stream members, because they are too faint to have good parallax. Finally, we used halo stars from where we identified streams. We found we underestimated distances of K-giants by 15\%, and overestimated distances of M-giants by 30\%, but no bias in BHBs, shown as left panels of Figure \ref{sys}. However, the systematic biases do not apply to Sgr stream members, because the difference between parallax and $1/d$ decreased with $G$ for both K-giants and M-giants, and most Sgr stream members are fainter than $G\sim15^{\rm{m}}$ (shown as right panels of Figure \ref{sys}).

Besides the distance $d$, our sample also includes equatorial coordinate information $(\alpha, \delta)$, heliocentric radial velocities $hrv$, and proper motions ($\mu_{\alpha}, \mu_{\delta}$). The $hrv$ of the LAMOST K-giants are obtained by the ULySS \citep{Wu11}, $hrv$ of LAMOST M-giants are calculated by \citet{Zhong19}, and $hrv$ of SEGUE K-giants and SDSS BHBs are from SEGUE Stellar Parameter Pipeline \citep[SPSS;][]{Lee08a, Lee08b}. The proper motions ($\mu_{\alpha}, \mu_{\delta}$) are from $Gaia$ DR2 by cross-matching with a radius of $1\arcsec$.

The chemical abundances (the overall metallicity [M/H] and the abundance of $\alpha$-element $[\alpha$/M]) of LAMOST K-, M-giants are from \citet{Zhang19}, which introduced a machine learning program called Stellar LAbel Machine (SLAM) to transfer the APOGEE DR15 \citep{Majewski17} stellar labels to LAMOST DR5 stars. The metallicity [Fe/H] of SDSS BHBs and SEGUE K-giants are estimated by SPSS. Since in the APOGEE data, [M/H] and [Fe/H] are calibrated using same method \citep{Holtzman15, Feuillet16}, we use [Fe/H] to represent the metallicity of all stars and do not to distinguish the [M/H] of LAMOST stars and [Fe/H] of SDSS/SEGUE stars hereafter.

For the measurement errors of our sample, LAMOST K-giants have a median distance precision of 13\%, a median radial velocity error of 7 km s$^{-1}$, a median error of 0.14 dex in metallicity, and a median $[\alpha$/Fe] error of 0.05 dex. SEGUE K-giants have a median distance precision of 16\% \citep{Xue14}, a median radial velocity error of 2 km s$^{-1}$, a typical error of 0.12 dex in metallicity. SDSS BHBs do not have error of individual star, but their distances are expected to be better than 10\% due to their nearly constant absolute magnitude \citep{Xue08}. The median radial velocity error of BHBs is 6 km s$^{-1}$, and the typical metallicity error is 0.22 dex. There is no distance error of individual LAMOST M-giant either. \citet{Li16} declared a typical distance precision of 20\%. LAMOST M-giants have a typical radial velocity error of about 5 km s$^{-1}$ \citep{Zhong15}, a median error of 0.17 dex in metallicity, and a median $[\alpha$/M] error of 0.06 dex. The proper motions of K-giants, M-giants, and BHBs are derived from $Gaia$ DR2, which is good to 0.2 mas yr$^{-1}$ at G = 17$^{\rm{m}}$.

Additionally, there are about 400 common K-giants between LAMOST and SEGUE samples, of which about 100 K-giants belong to Sgr streams. We used these common K-giants to find that LAMOST K-giants have a $-$8.1 km s$^{-1}$ offset in radial velocity from SEGUE K-giants, and the two surveys have consistent metallicities and distance. In this paper, we have added 8.1 km s$^{-1}$ to LAMOST K-giants to avoid systematic bias from SEGUE K-giants. In analysis of Sgr streams, the duplicate K-giants are removed. See Table \ref{t_catalog} for an example of the measurements and corresponding uncertainties.

\subsection{Integrals of Motion and Friends-of-Friends Algorithm}
To search stars with similar orbits through friends-of-friends (FoF), X19 defined five IoM parameters: eccentricity $e$, semimajor axis $a$, direction of the orbital pole $(l_{\rm{orb}}, b_{\rm{orb}})$, and the angle between apocenter and the projection of $X$-axis on the orbital plane $l_{\rm{apo}}$. Then they calculated the ``distance" between any two stars in the normalized space of $(e, a, l_{\rm{orb}}, b_{\rm{orb}}, l_{\rm{apo}})$ and used FoF to find out the group stars that have similar orbits according to the size of the ``distance". The five IoM parameters $(e,a,l_{\rm{orb}}, b_{\rm{orb}}, l_{\rm{apo}})$ are gotten by 6D information $(\alpha, \delta,d,hrv,\mu_{\alpha}, \mu_{\delta})$ under the assumption that the Galactic potential is composed of a spherical \citet{Hernquist90} bulge, an exponential disk, and a NFW halo \citep{NFW96}. See Table \ref{t_orbs} for an example of the orbital parameters and corresponding uncertainties.

By comparing the FoF groups with observations and simulations of Sgr \citep{LM10, Koposov12, Belokurov14, DL17, Hernitschek17}, X19 identified 3028 Sgr stream members, including 2626 K-giants (including 102 suspected duplicate stars), 158 M-giants, and 224 BHBs, which is the largest spectroscopic sample obtained in the Sgr stream yet. In the next section, we will exhibit the Sgr members in detail, including spatial, kinematic, and abundance features.

\section{THE PROPERTIES of SAGITTARIUS STREAM}\label{sec:Sgr}
The Cartesian reference frame used in this work is centered at the Galactic center, the $X$-axis is positive toward the Galactic center, the $Y$-axis is along the rotation of the disk, and the $Z$-axis points toward the North Galactic Pole. We adopt the Sun's position is at $(-8.3,0,0)$ kpc \citep{de_Grijs16}, the local standard of rest (LSR) velocity is 225 km s$^{-1}$ \citep{de_Grijs17}, and the solar motion is $(+11.1,+12.24,+7.25)$ km s$^{-1}$ \citep{Schonrich10}.

\begin{figure}[htb]
\centering
  \begin{tabular}{@{}cccc@{}}
    \includegraphics[width=.9\textwidth]{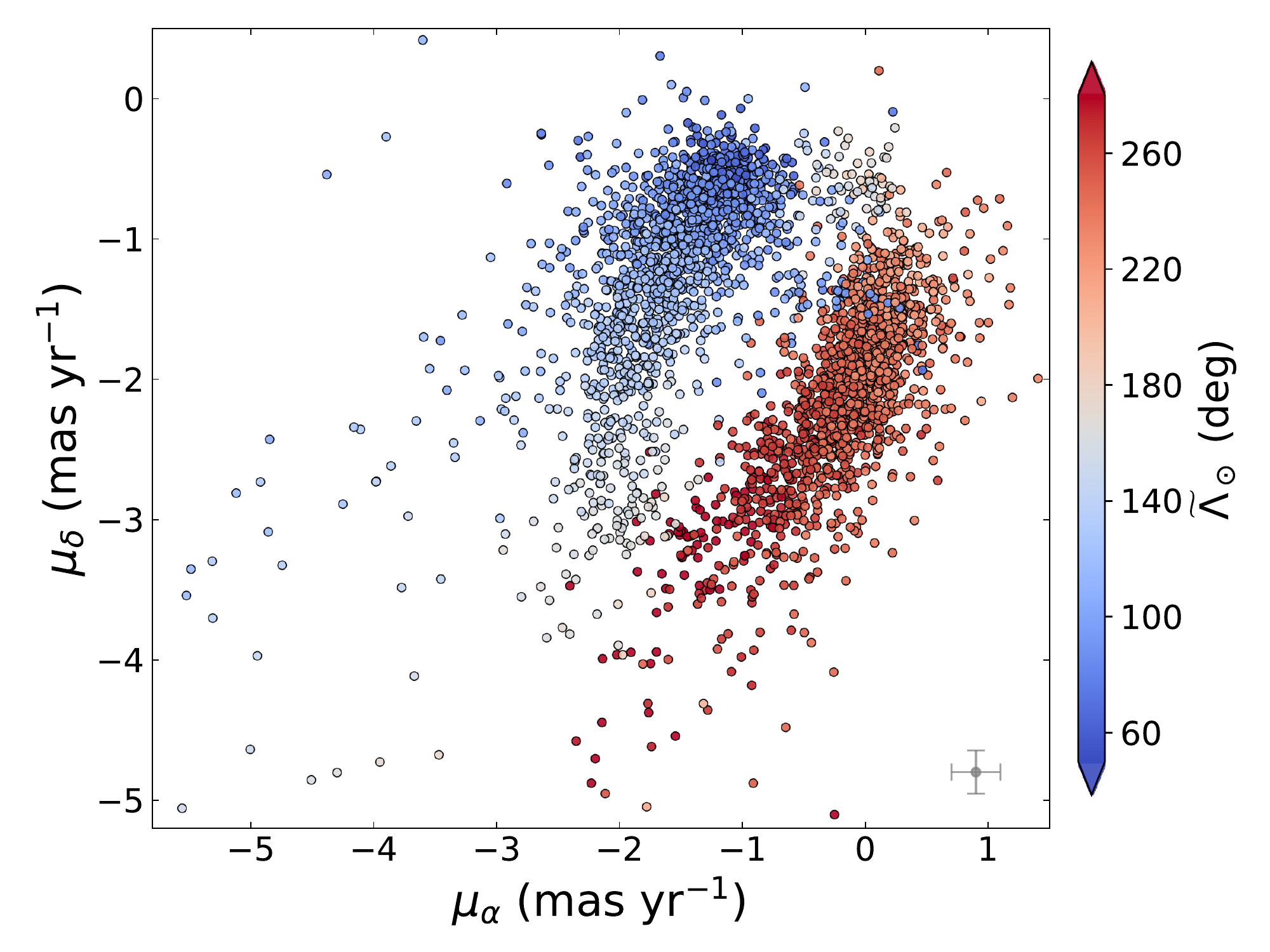}
  \end{tabular}
\caption{Distribution of Sgr stream in proper motions ($\mu_{\alpha}, \mu_{\delta}$) space and color-coded by $\widetilde{\Lambda}_\odot$. The colors can help to identify the Sgr leading (blue cluster) and trailing stream (red cluster). A mean error bar is shown in the lower right corner.
} \label{pm}
\end{figure}

Figure \ref{pm} presents the proper motions ($\mu_{\alpha}, \mu_{\delta}$) of Sgr stream stars. The colors represent the longitude in Sgr coordinate system, $\widetilde{\Lambda}_\odot$, and help to identify the stars belonging to different Sgr streams. In this figure, we can easily see the variation of proper motion along the leading and trailing stream.

\begin{figure}[htb]
\centering
  \begin{tabular}{@{}cccc@{}}
    \includegraphics[width=.85\textwidth]{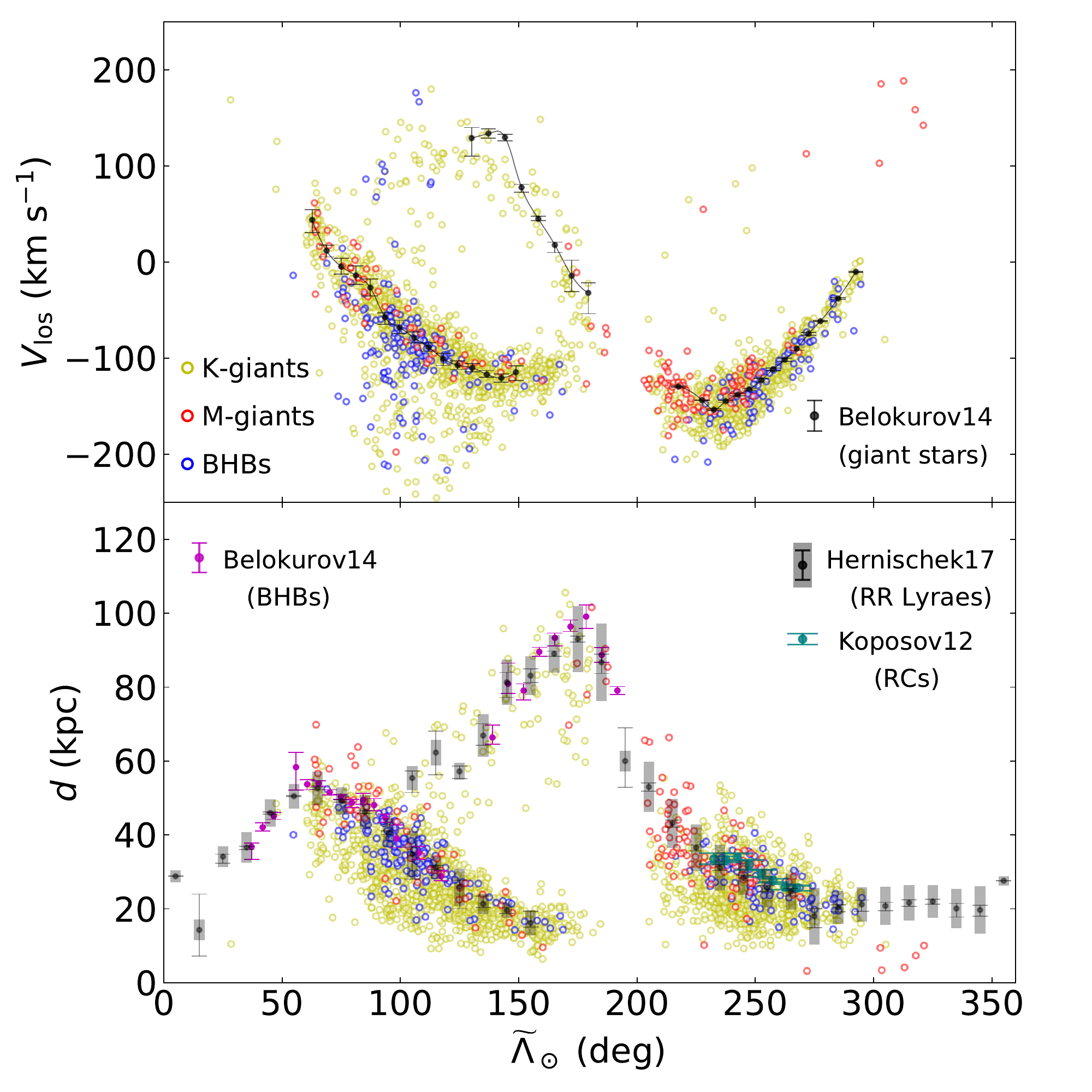} &
  \end{tabular}
\caption{Comparisons with observed Sgr data in coordinates of $(\widetilde{\Lambda}_\odot, d)$ and $(\widetilde{\Lambda}_\odot, V_{\rm{los}})$. Yellow, red, and blue circles represent K-, M-giants, and BHBs separately. In the top panel, the black dots with error bars from tables 3-5 of \citet{Belokurov14}, which estimated by their Sgr giant stars. In the bottom panel, magenta and green dots represent BHBs and RCs from tables 1-2 of \citet{Belokurov14} and table 2 of \citet{Koposov12}, the black dots are from tables 4-5 of \citet{Hernitschek17} obtained from RR Lyrae stars. The error bars mean distance uncertainty, and the gray bars represent 1$\sigma$ line-of-sight depth of the Sgr stream.
} \label{obs}
\end{figure}

\begin{figure}[htb]
\centering
  \begin{tabular}{@{\hspace{-1.0cm}}cccc@{}}
    \includegraphics[width=1.1\textwidth]{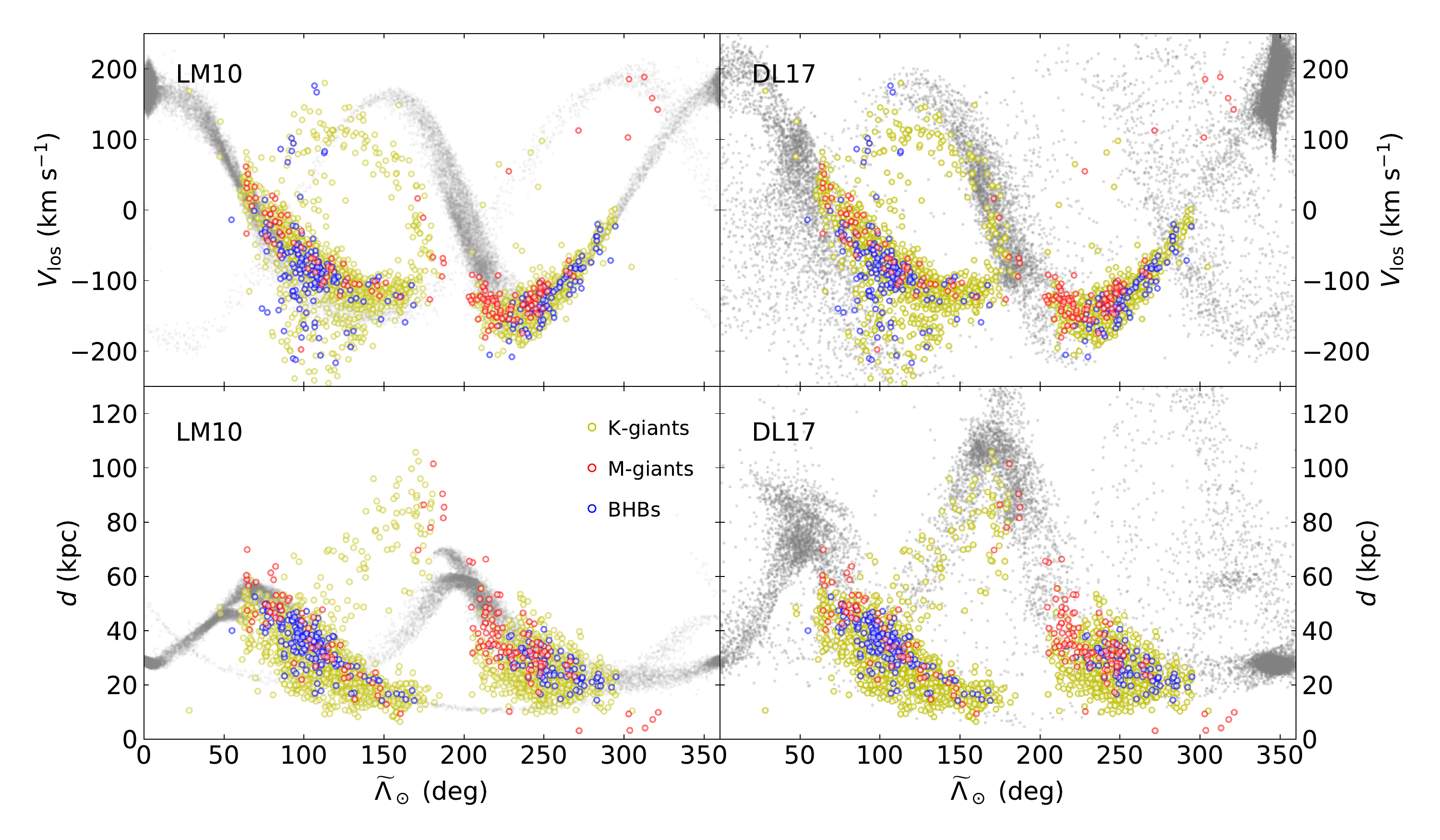} &
  \end{tabular}
\caption{Comparisons with Sgr simulations in coordinates of $(\widetilde{\Lambda}_\odot, d)$ and $(\widetilde{\Lambda}_\odot, V_{\rm{los}})$. Grey dots in the left and right panels respectively indicate the model of \citetalias{LM10} and \citetalias{DL17}. Yellow, red, and blue circles separately represent K-, M-giants, and BHBs.
} \label{sims}
\end{figure}

Figure \ref{obs} shows the Sgr streams traced by K-giants, M-giants and BHBs are consistent with previous observations both in line-of-sight velocities \citet{Belokurov14} and distances \citep{Koposov12, Belokurov14, Hernitschek17}. The comparison with simulations is presented in Figure \ref{sims}. In the range of $100^\circ < \widetilde{\Lambda}_\odot < 200^\circ$ and $d > 60$ kpc, both velocities and distances do not match with \citet{LM10} (\citetalias{LM10}) model shown as left panel of Figure \ref{sims}. The right panel of Figure 3 shows Sgr streams traced by K-giants, M-giants and BHBs are roughly in good agreement with \citet{DL17} (\citetalias{DL17}) both in velocities and distances. In the range of $\sim100^\circ < \widetilde{\Lambda}_\odot < 150^\circ$ and $V_{\rm{los}}$ 130 km s$^{-1}$, the observation shows slightly slow than \citetalias{DL17} simulation. Furthermore, we have fewer stars beyond 100 kpc than the prediction of \citetalias{DL17}, which we attribute to the limiting magnitude of LAMOST (r $\sim17.8^{\rm{m}}$). On the Sgr orbital plane, $\widetilde{\Lambda}_\odot$ $< 50^\circ$ and $\widetilde{\Lambda}_\odot$ $> 300^\circ$ is out of the Sky coverage of LAMOST and SDSS/SEGUE, where is around the Sgr dSph.

\subsection{Kinematics of Sagittarius Stream}
\begin{figure}[htb]
\centering
  \begin{tabular}{@{}cccc@{}}
    \includegraphics[width=.95\textwidth]{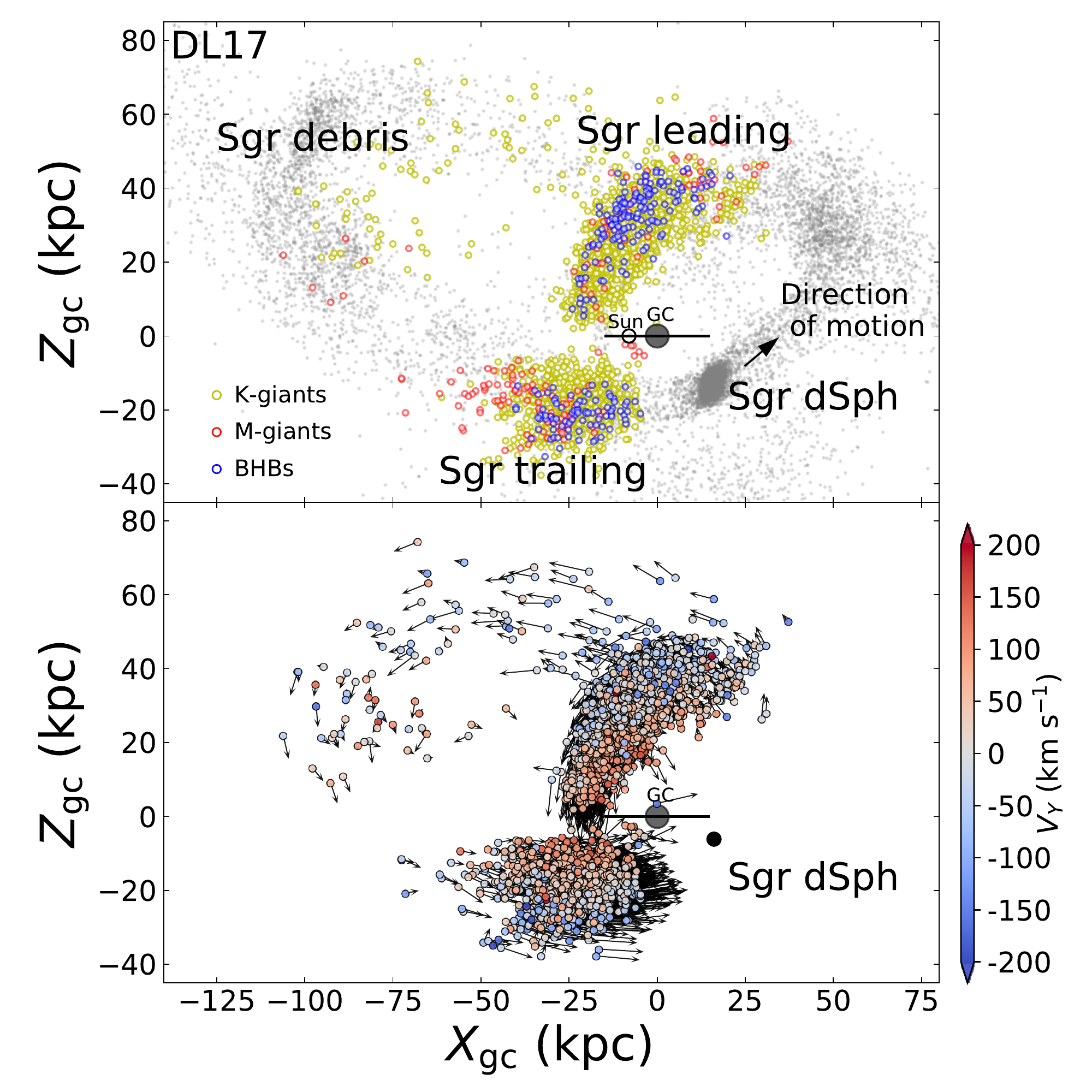} &
  \end{tabular}
\caption{Top panel: distribution of each Sgr component (Sgr dSph, Sgr leading, Sgr trailing, and Sgr debris) in $X$-$Z$ plane. Yellow, red, and blue circles represent the K-, M-giants, and BHBs, and grey dots indicate the \citetalias{DL17} model. Bottom panel: arrows represent the Sgr stars' moving directions and velocity amplitudes, and every star is color-coded according to its velocity along $Y$-axis ($V_Y$). Because of the coverage limits of LAMOST and SDSS/SEGUE, we have no Sgr stars around the Sgr dSph.} \label{xz}
\end{figure}

Figure \ref{xz} illustrates the spatial distribution of Sgr in $X$-$Z$ plane, which is close to the Sgr's orbital plane. In the top panel, we show our Sgr sample with \citetalias{DL17} model as background. We tag the position of each Sgr component (Sgr dSph, Sgr leading, Sgr trailing, and Sgr debris), and Sgr dSph's moving direction. The panel exhibits the position of each Sgr component in spatial distribution, and our sample comports with \citetalias{DL17} model perfectly. In the bottom panel, the arrows indicate the direction and amplitude of velocities in $X$-$Z$ plane and every star is color-coded according to its velocity component in and out of $X$-$Z$ plane ($V_{Y}$). This panel well illustrates the kinematic feature of stream, i.e., stream stars move together in phase space. Besides, the arrows and low latitude M-giants (red circles in the top panel) implies that the Sgr debris actually is the continuation of the Sgr trailing stream and where the trailing stream stars return from their apocenter. Thus, the apogalacticon of Sgr trailing stream could reach $\sim100$ kpc from the Sun (see $\widetilde{\Lambda}_\odot \sim170^\circ$ in Figure \ref{obs}). This apogalacticon is consistent with the work of \citet{Belokurov14}, \citet{Koposov15}, \citet{Sesarl17}, \citet{Hernitschek17}, and \citet{Li19}. In addition, the panel also presents a obvious gradient in $V_Y$ along the line-of-sight direction from the Galactic center in both leading and trailing stream.

\begin{figure}[htb]
\centering
  \begin{tabular}{@{\hspace{-0.5cm}}cccc@{}}
    \includegraphics[width=.5\textwidth]{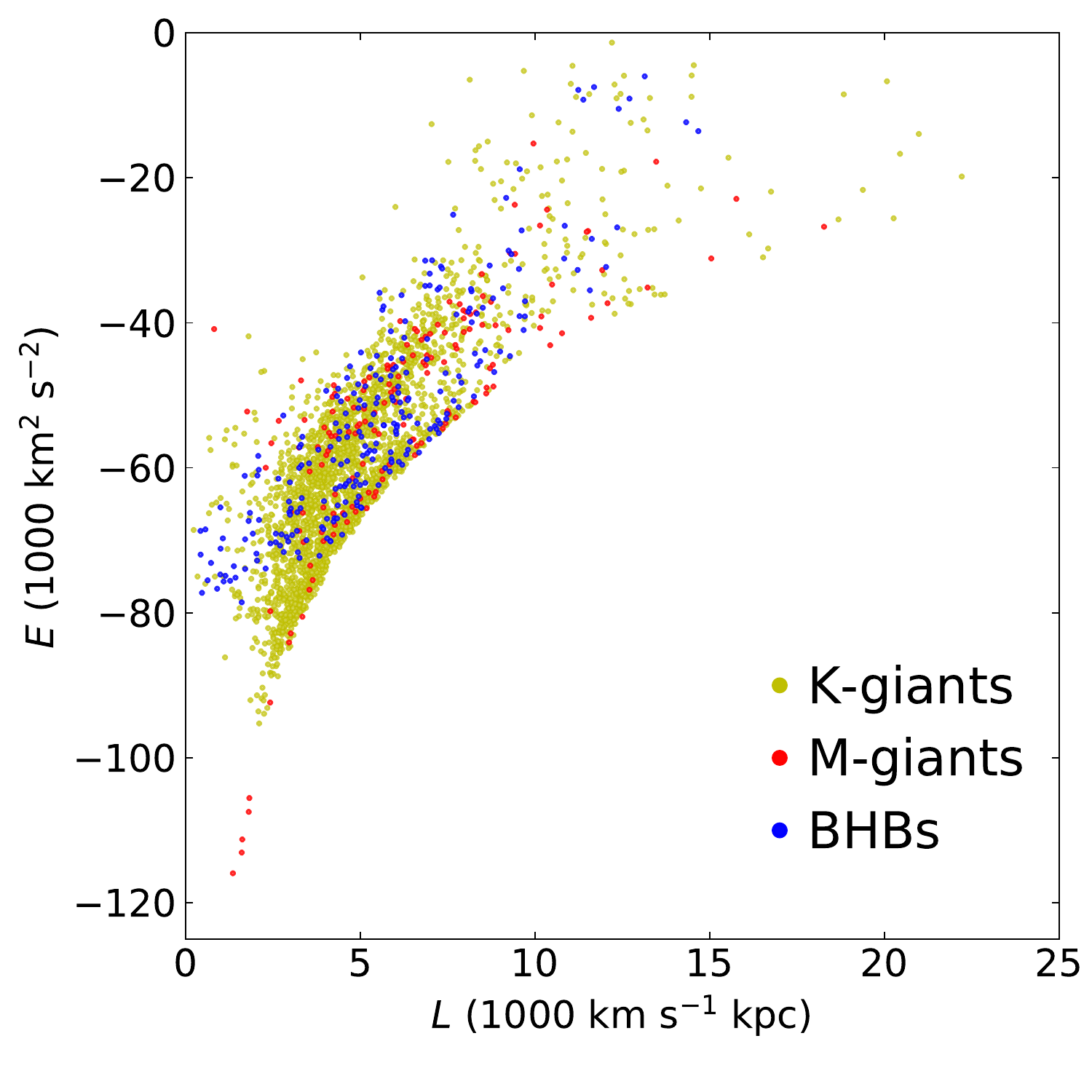} &
    \includegraphics[width=.5\textwidth]{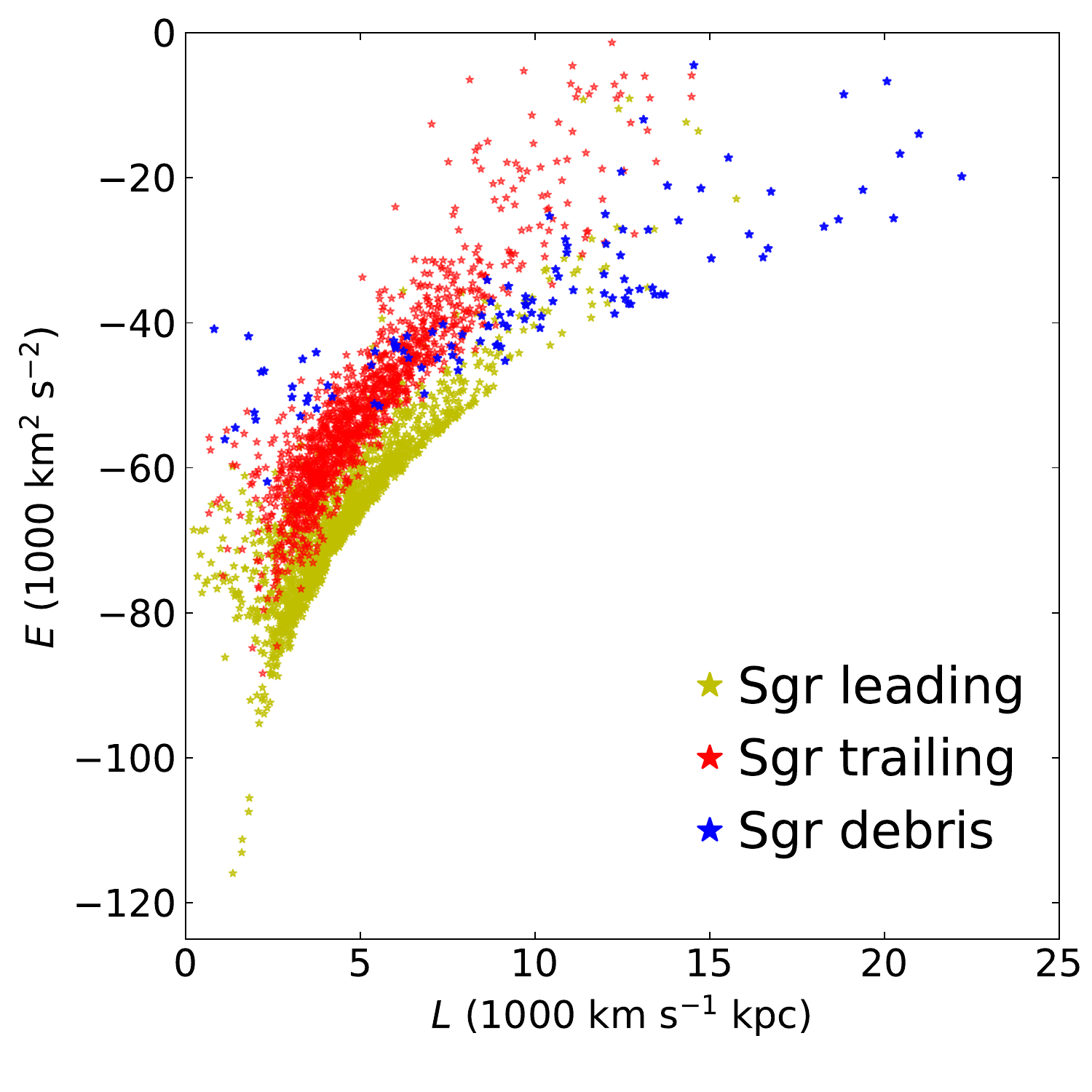} &\\
  \end{tabular}
\caption{Sgr stars in angular momentum ($L$) versus energy ($E$) space. Left and right panel respectively exhibit the distribution of different kinds of stars and streams in $E$-$L$ space.
} \label{le}
\end{figure}

In Figure \ref{le}, we examine the angular momentum ($L$) and energy ($E$) of Sgr member stars. The left panel shows the Sgr K-, M-giants and BHBs in $E$-$L$ space, and there is no tangible difference among them. The right panel illustrates the stars from different Sgr streams. The panel shows the energy of each stream are quite different, Sgr debris and trailing stream are significantly higher than leading stream.

\subsection{Metallicities of Sagittarius Stream}\label{sub_sec:feh_vy}
\begin{figure}[htb]
\centering
  \begin{tabular}{@{}cccc@{}}
    \includegraphics[width=.5\textwidth]{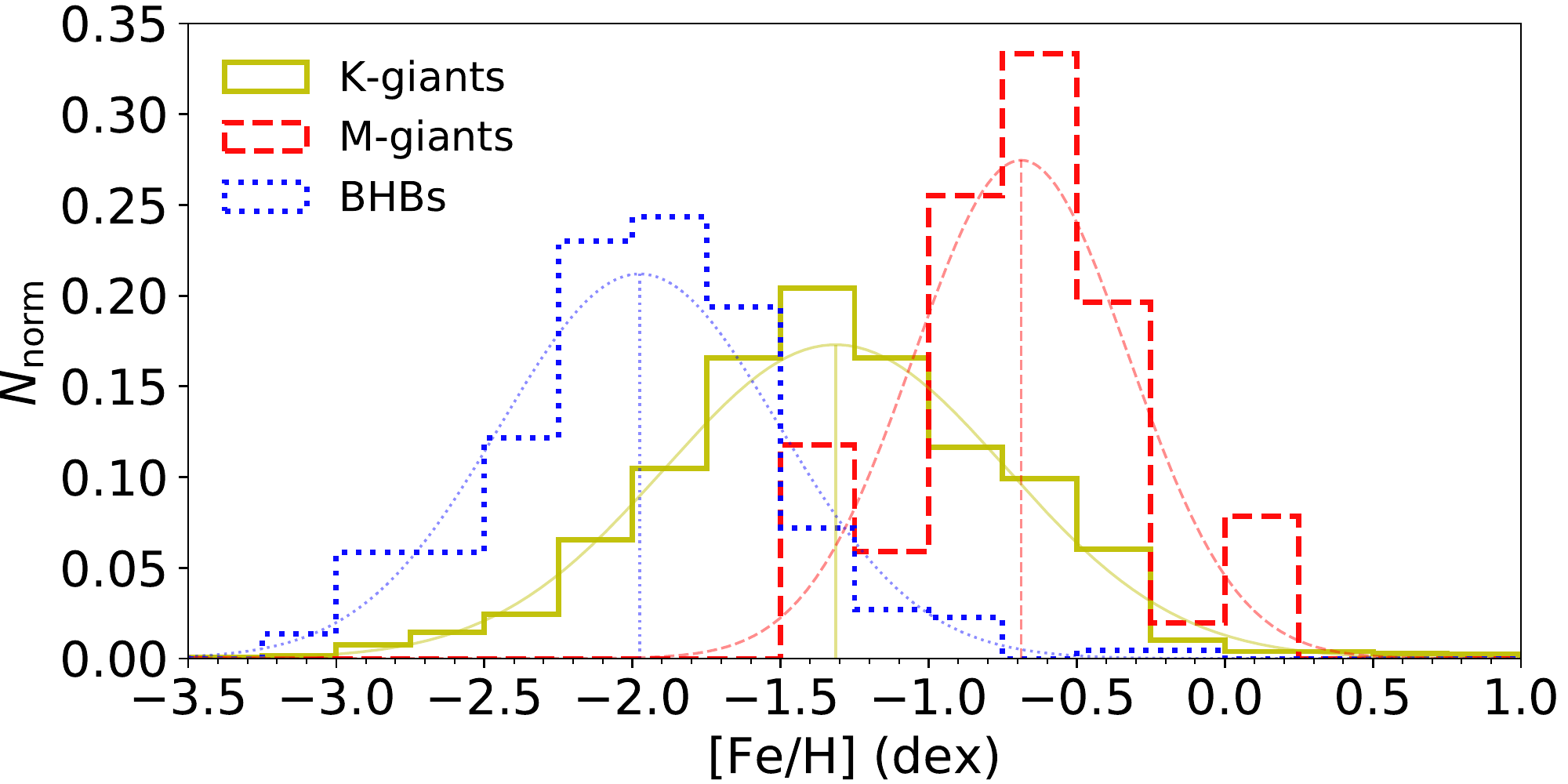} &
    \includegraphics[width=.5\textwidth]{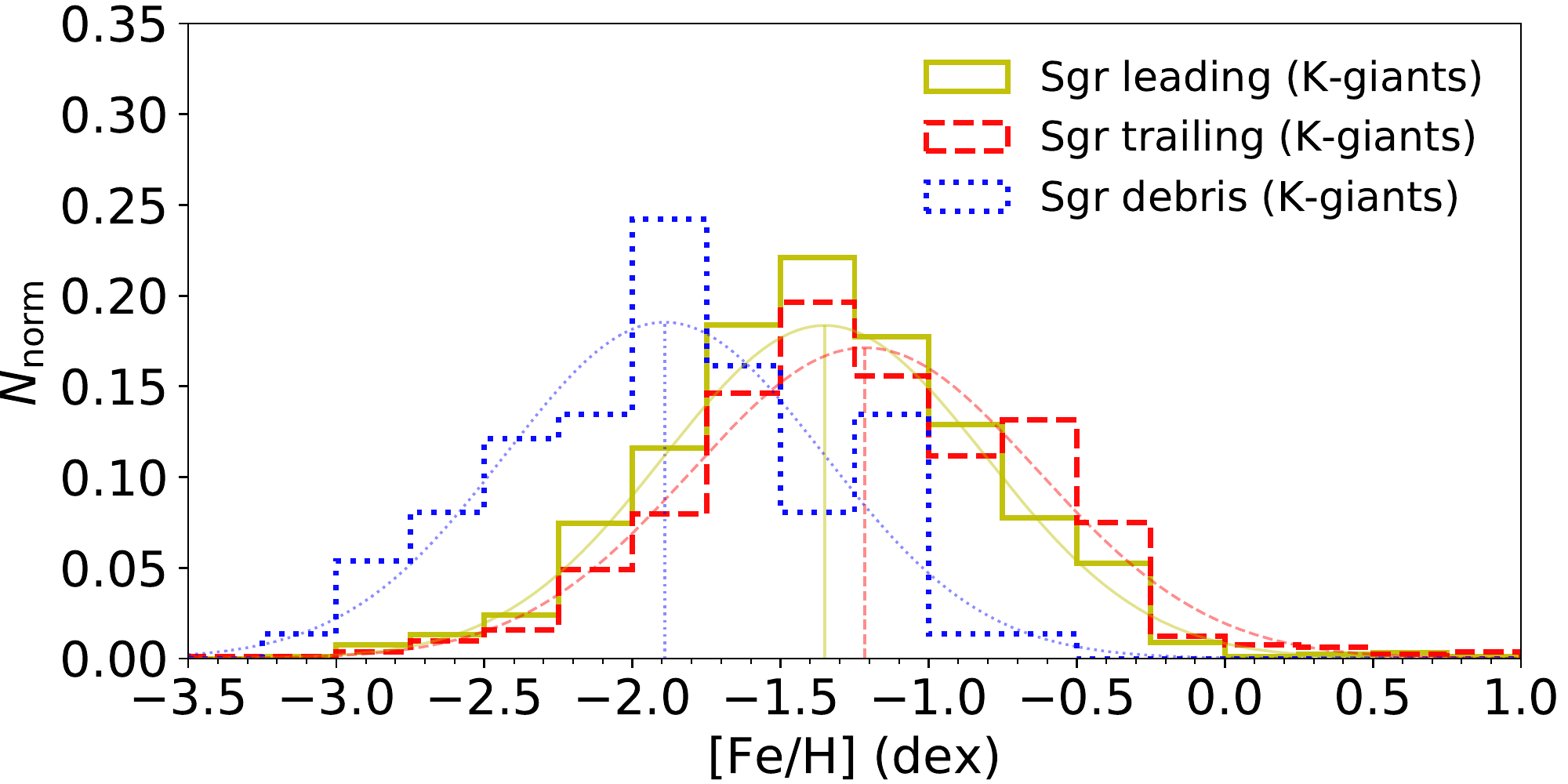} &\\
    \includegraphics[width=.5\textwidth]{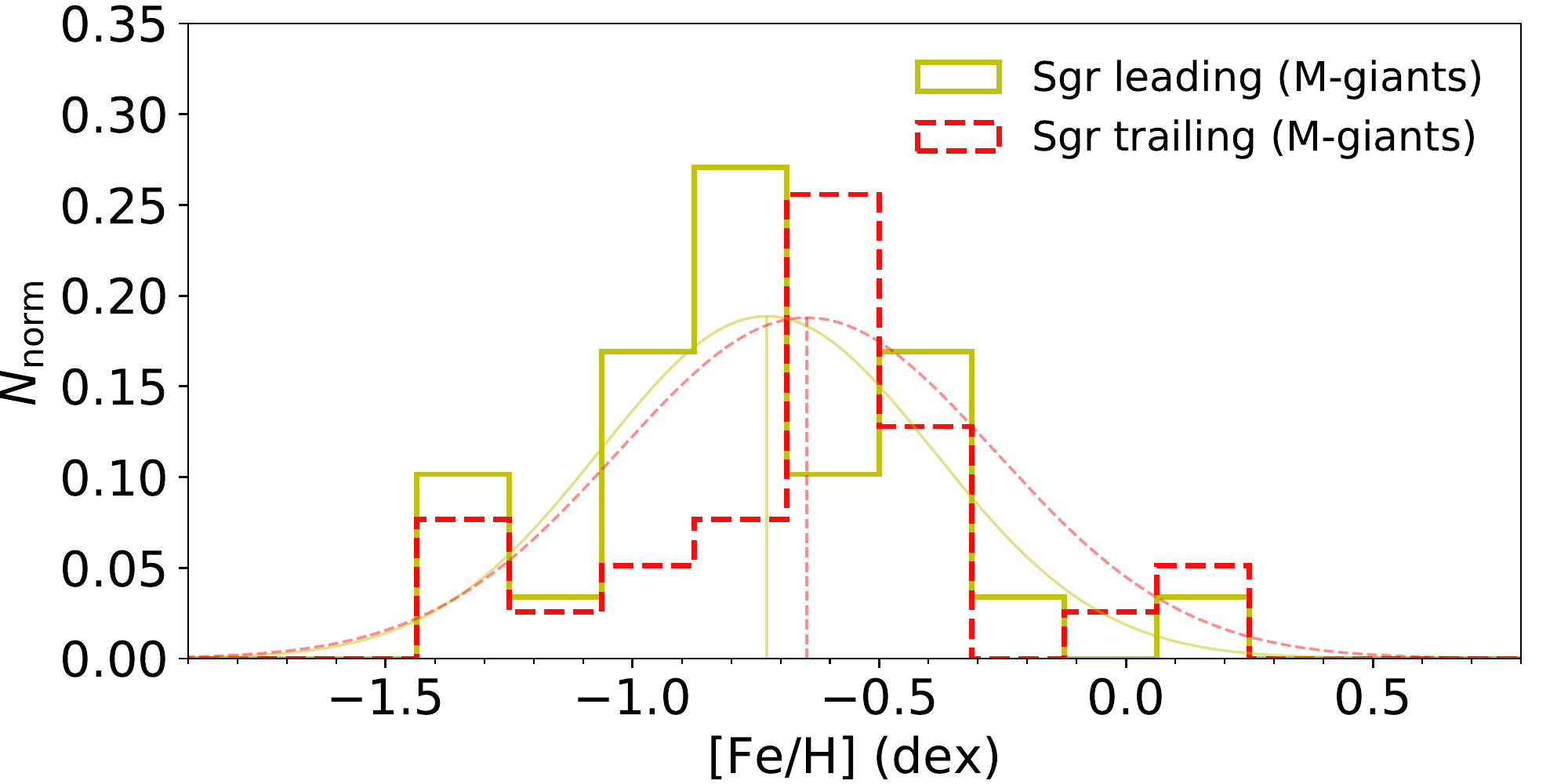} &
    \includegraphics[width=.5\textwidth]{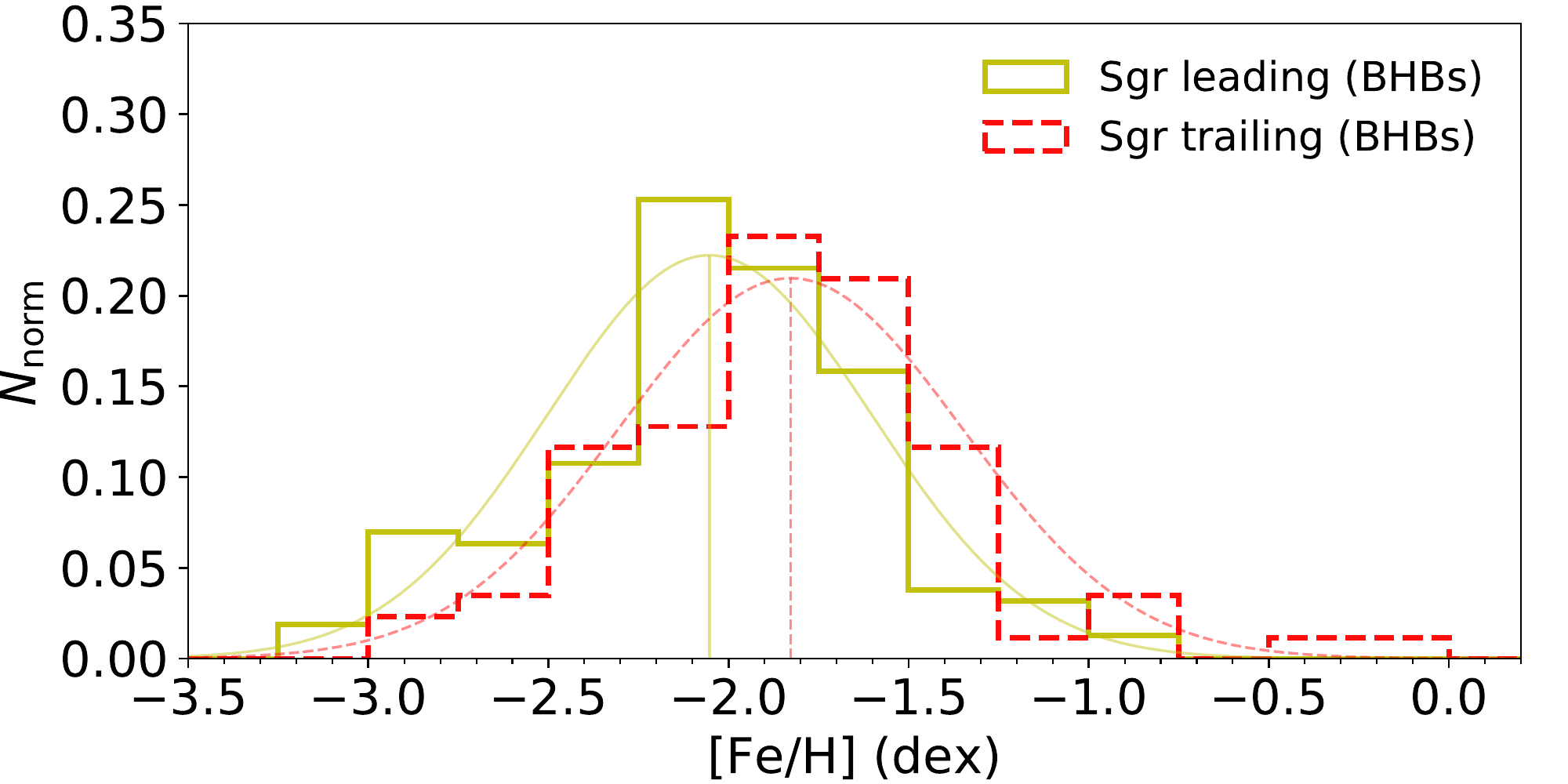} &\\
  \end{tabular}
\caption{The metallicity distribution of Sgr K-, M-giants and BHBs (top left panel) and different streams (top right and bottom panels). Histogram with different colors represent different kinds of stars or streams, and each histogram has corresponding gaussian distribution obtained by the mean metallicity $<\rm{[Fe/H]}>$ and scatter $\sigma_{\rm{[Fe/H]}}$. In the bottom panels, there are few K-giants and BHBs can be used to plot histogram of Sgr debris.
} \label{feh_hist}
\end{figure}

Figure \ref{feh_hist} presents the metallicity distribution of our Sgr sample. In the top left panel, we exhibit the sample's metallicity distribution from K-, M-giants, and BHBs. The panel shows that M-giants is the most metal-rich population with mean metallicity $<\rm{[Fe/H]}>$ = $-0.69$ dex and scatter $\sigma_{\rm{[Fe/H]}}$ = 0.36 dex, BHBs is the most metal-poor population with $<\rm{[Fe/H]}>$ = $-1.98$ dex and $\sigma_{\rm{[Fe/H]}}$ = 0.47 dex, and for the K-giants, these values are [Fe/H] = $-1.31$ dex and $\sigma_{\rm{[Fe/H]}}$ = 0.58 dex. The mean metallicity of Sgr M-giants is close to the result of high-resolution spectra from \citet{Carlin18}, which used 42 Sgr stream common stars of LAMOST DR3 M-giants ($-0.68$ dex for trailing stream and $-0.89$ dex for leading stream). This implies that the metallicity of our LAMOST sample is reliable. In the other panels, we pick up K-, M-giants, and BHBs to exhibit the metallicity of Sgr leading, trailing, and debris separately. The top right panel (K-giants) shows that the Sgr leading stream has $<\rm{[Fe/H]}>$ = $-1.35$ dex with $\sigma_{\rm{[Fe/H]}}$ = 0.54 dex, the Sgr trailing stream has $<\rm{[Fe/H]}>$ = $-1.21$ dex and $\sigma_{\rm{[Fe/H]}}$ = 0.58 dex, and for Sgr debris, $<\rm{[Fe/H]}>$ = $-1.89$ dex and $\sigma_{\rm{[Fe/H]}}$ = 0.54 dex. Thus, Sgr trailing stream is on average the most metal-rich Sgr stream, followed by Sgr leading and debris. In bottom panels, the M-giants and BHBs present a similar feature, i.e., the trailing stream is more metal-rich than leading stream. This difference among Sgr different streams had been mentioned in \citet{Carlin18}, and they suggested that this difference might cause by the stars' different unbound time from Sgr core.

\begin{figure}[htb]
\centering
  \begin{tabular}{@{\hspace{-1.0cm}}cccc@{}}
    \includegraphics[width=.55\textwidth]{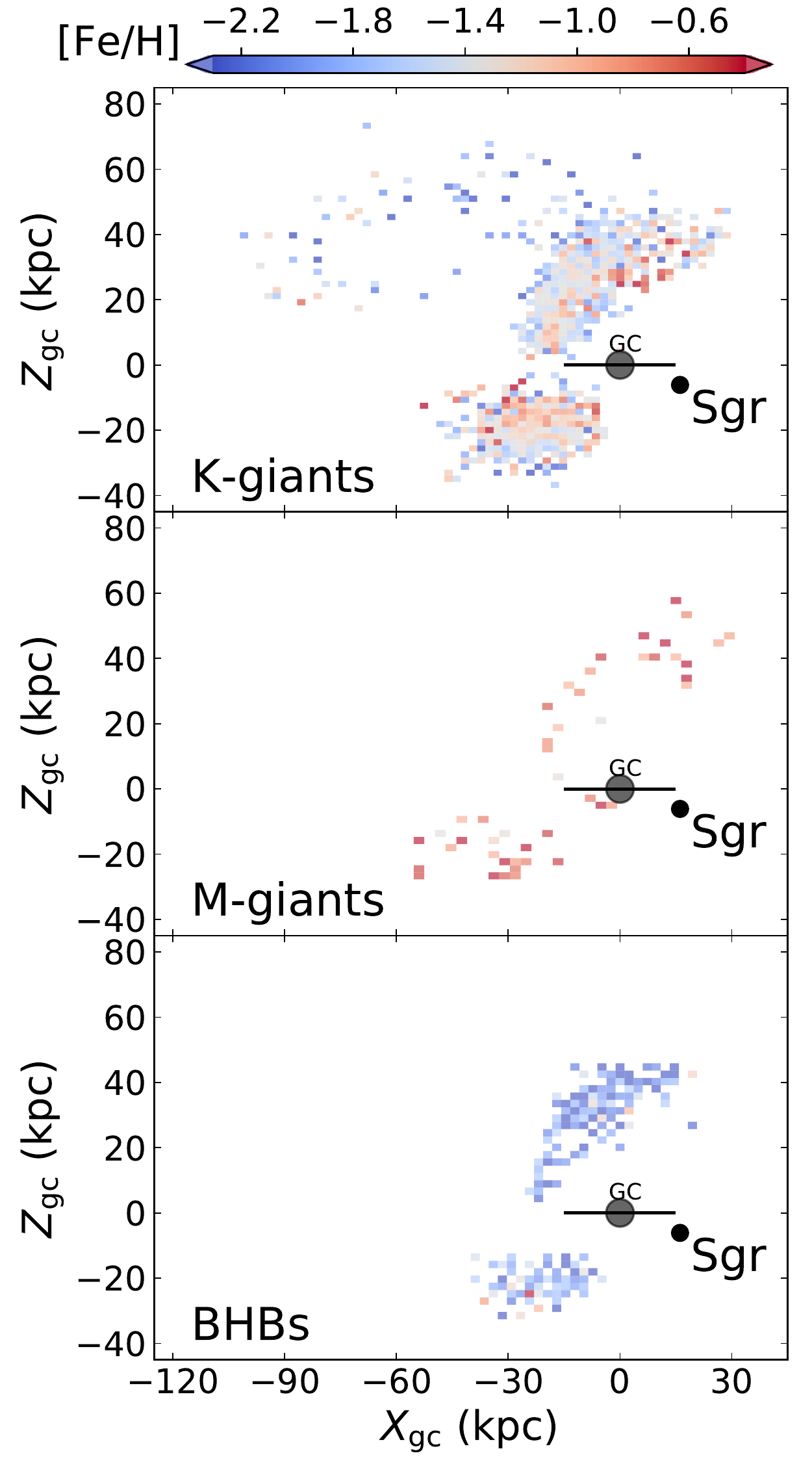} &
    \includegraphics[width=.55\textwidth]{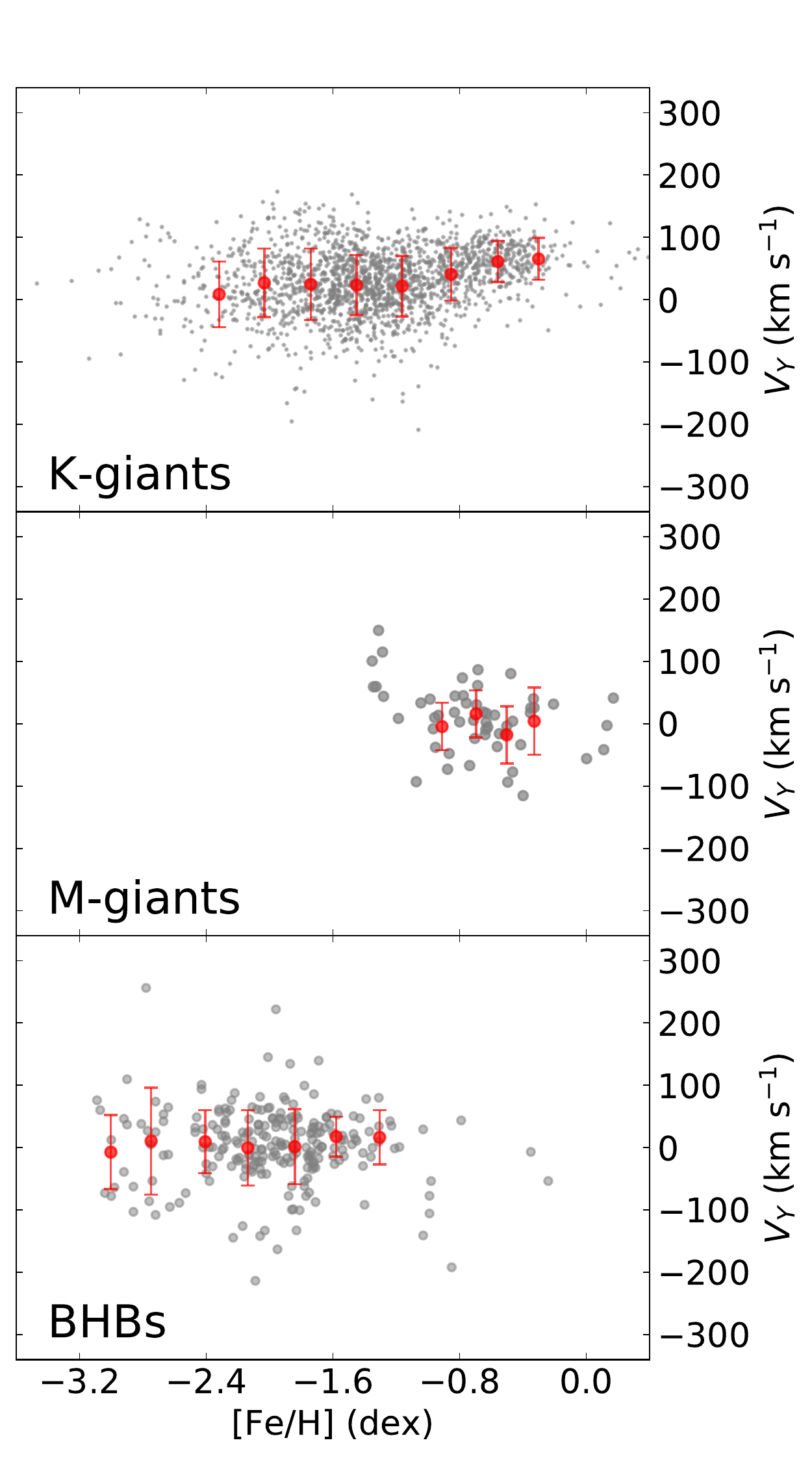} &
    \\
  \end{tabular}
\caption{Left panels: the metallicity distribution of Sgr K-, M-giants and BHBs in $X$-$Z$ plane. Right panels: distribution of Sgr K-, M-giants and BHBs in [Fe/H] versus $V_Y$ space. Red dots with error bars represent the mean of $V_Y$ and its dispersion.
} \label{feh_xz}
\end{figure}

In Figure \ref{feh_xz}, we present the metallicity distribution of Sgr stars in $X$-$Z$ plane. Similar with Figure \ref{xz}, the K-giants in the top left panel shows that the metallicity also has a gradient along the line-of-sight direction, which indicates that the inner side stars (close to the Galactic center) are not only different with outer side stars (away from the Galactic center) in kinematics ($V_Y$), but also in metallicity. In the top right panel, we plot the K-giants in the [Fe/H] versus $V_Y$ space. The panel shows that $V_Y$ increases with metallicity at [Fe/H] $>\sim -$1.5 dex, which implies that there are some correlations between $V_Y$ and metallicity in Sgr stream. In the distribution of M-giants and BHBs, we do not see clear feature as K-giants have.

\subsection{Alpha-Abundances of Sagittarius Stream}
\begin{figure}[htb]
\centering
  \begin{tabular}{@{\hspace{-0.2cm}}c@{}}
  \includegraphics[width=0.9\textwidth]{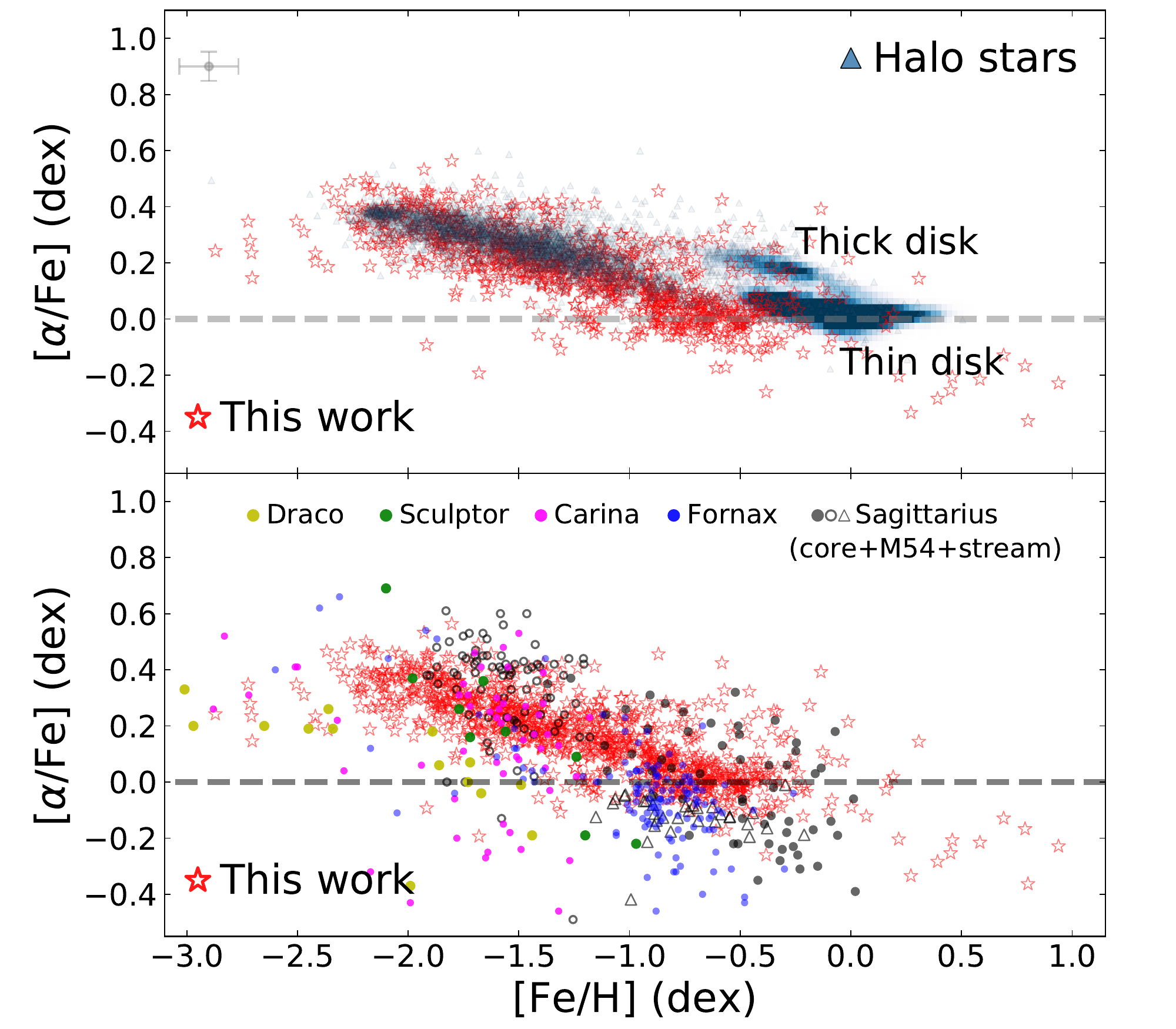}
  \end{tabular}
\caption{Comparisons of LAMOST Sgr stars with the Milky Way components (upper panel) and dwarf galaxies (lower panel) in [$\alpha$/Fe] versus [Fe/H] space. In upper panel, we plot red stars as our Sgr sample, blue density map as disk stars, and blue triangles as halo stars. The disk stars are selected by $|Z|<3$ kpc, and for halo stars, we choose $|Z|>$ 5 kpc and not belonging to any substructures. A typical error bar of our Sgr stars is presented on top-left corner. In the lower panel, our Sgr stars are shown as red stars, and the Milky Way dwarf galaxies including Draco \citep[yellow dots;][]{Shetrone01, CH09}, Sculptor \citep[green dots;][]{Shetrone03, Geisler05}; Carina \citep[magenta dots;][]{Koch08, Lemasle12, Shetrone03, Venn12}; Fornax \citep[blue dots;][]{Letarte10, Lemasle14}; Sagittariuis core (black dots), M54 (black circles), and stream (black triangles) \citep{Monaco05, Sbordone07, Carretta10, McWilliam13, Hasselquist19}).
} \label{feh_alpha}
\end{figure}

It is well established that dwarf galaxies have different chemical-evolution paths with the Milky Way \citep{Tolstoy09, Kirby11}. In Figure \ref{feh_alpha}, we present the abundance of $\alpha$-element from LAMOST Sgr stars obtained by SLAM \citep{Zhang19}. In top panel, we compare the Sgr sample with the Milky Way stars, including the Galactic disk and halo. For disk, we choose stars with $|Z|<$ 3 kpc (blue density map), and for halo, we plot the stars with $|Z|>$ 5 kpc and not belonging to any substructures (blue dots; X19). The top panel shows that the trend of [$\alpha$/Fe] is similar with halo stars at lower metallicity, but the ratio then evolve down to lower values than disk stars at higher metallicity. In addition, there might be a hint of a knee at [Fe/H] $\sim -2.3$ dex, but it is not very clear in our data. If the knee is very metal-poor (or non-existent), then Sgr must have had a very low star-formation efficiency at early times (similar to, e.g., the Large Magellanic Cloud; \citealt{Nidever19}). In the bottom panel, we compare the $\alpha$-abundance ([Mg/Fe]) with previous work of Sgr, including M54 \citep{Carretta10}, Sgr core \citep{Monaco05, Sbordone07, Carretta10, McWilliam13}, and Sgr stream \citep{Hasselquist19}. In the panel, our Sgr stream sample mainly follows the stars in M54 and Sgr core, but are slightly higher in $\alpha$-abundance than the Sgr stream stars from \citet{Hasselquist19} in the same range of metallicity  ($-1.2 < {\rm[Fe/H]} < -0.2$ dex). We also include [Mg/Fe] versus [Fe/H] of some other dwarf galaxies, like Draco \citep{Shetrone01, CH09}, Sculptor \citep{Shetrone03, Geisler05}, Carina \citep{Koch08, Lemasle12, Shetrone03, Venn12}, Fornax \citep{Letarte10, Lemasle14}, and the panel shows a similar evolution pattern of $\alpha$-element between our Sgr stream and dwarf galaxies.

\subsection{Bifurcations in Sagittarius Stream}
\begin{figure}[htb]
\centering
  \begin{tabular}{@{\hspace{-0.2cm}}c@{}}
  \includegraphics[width=1.1\textwidth]{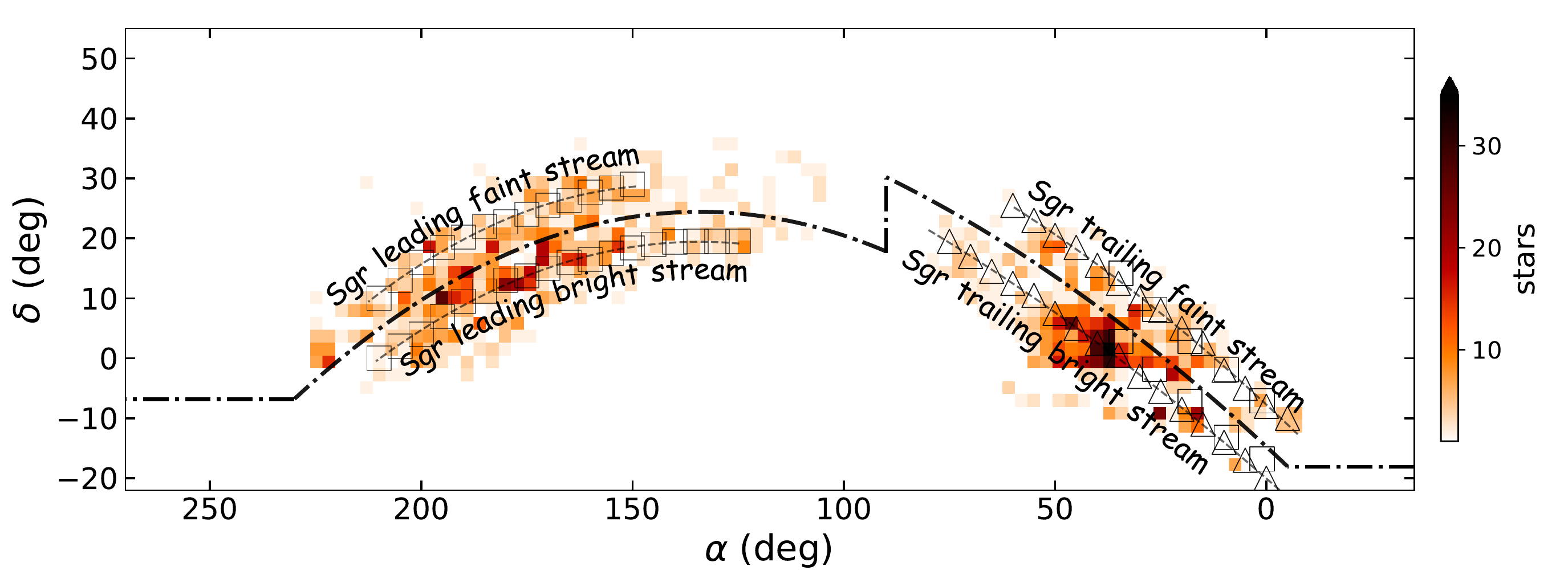}
  \end{tabular}
\caption{Density map of Sgr bifurcation in equatorial coordinate. The squares and dashed lines are defined in \citet{Belokurov06} and \citet{Koposov12}, the triangles are defined in Table \ref{t_bif}, and the dash-dotted line between faint and bright streams is used to distinguish faint and bright stream stars.
} \label{bif_div}
\end{figure}

In Figure \ref{bif_div}, we exhibit the Sgr bifurcation in density map using our sample. To identify the faint and bright stream of the bifurcation, we add the coordinates of faint and bright stream defined by \citet{Belokurov06} and \citet{Koposov12} (see squares in Figure \ref{bif_div}). In addition, we extend the coordinates of Sgr bifurcation in trailing stream based on our sample (see Table \ref{t_bif} and the triangles in Figure \ref{bif_div}). The dash-dotted line between faint and bright stream is used to distinguish the faint and bright stream stars, above the dash-dotted line are belonging to faint stream, and below are belonging to bright stream. In previous studies, the Sgr bifurcation was identified through density map from photometry data \citep{Koposov12, Belokurov14}, and it is called bright stream because it is denser than faint stream in density map. Due to few spectroscopic data of Sgr streams, it is hard to statistically analyze the kinematics and chemistry of the bifurcation. LAMOST and SEGUE sample provided many spectra of Sgr stream stars in either bright or faint stream, which allow us to analyze the properties of the bifurcation in detail.

\begin{figure}[htb]
\centering
  \begin{tabular}{@{\hspace{-0.2cm}}c@{}}
  \includegraphics[width=1.1\textwidth]{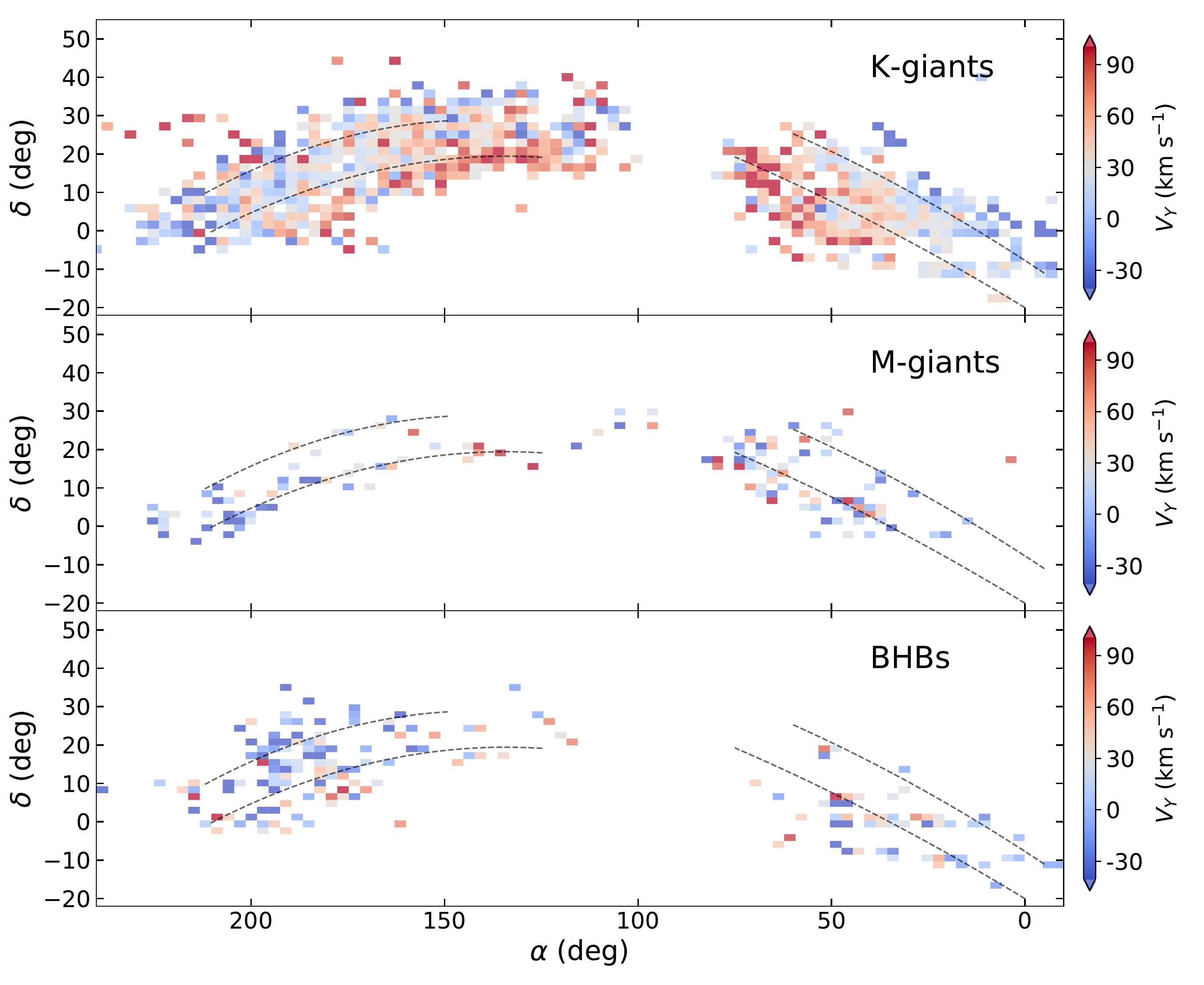}
  \end{tabular}
\caption{Density map of Sgr bifurcation from K-, M-giants and BHBs in equatorial coordinate. Colors represents the velocity along $Y$-axis $V_Y$. Dashed lines represent the center line of the faint and bright streams.
} \label{bif_vy}
\end{figure}

\begin{figure}[htb]
\centering
  \begin{tabular}{@{\hspace{-0.2cm}}c@{}}
  \includegraphics[width=1.1\textwidth]{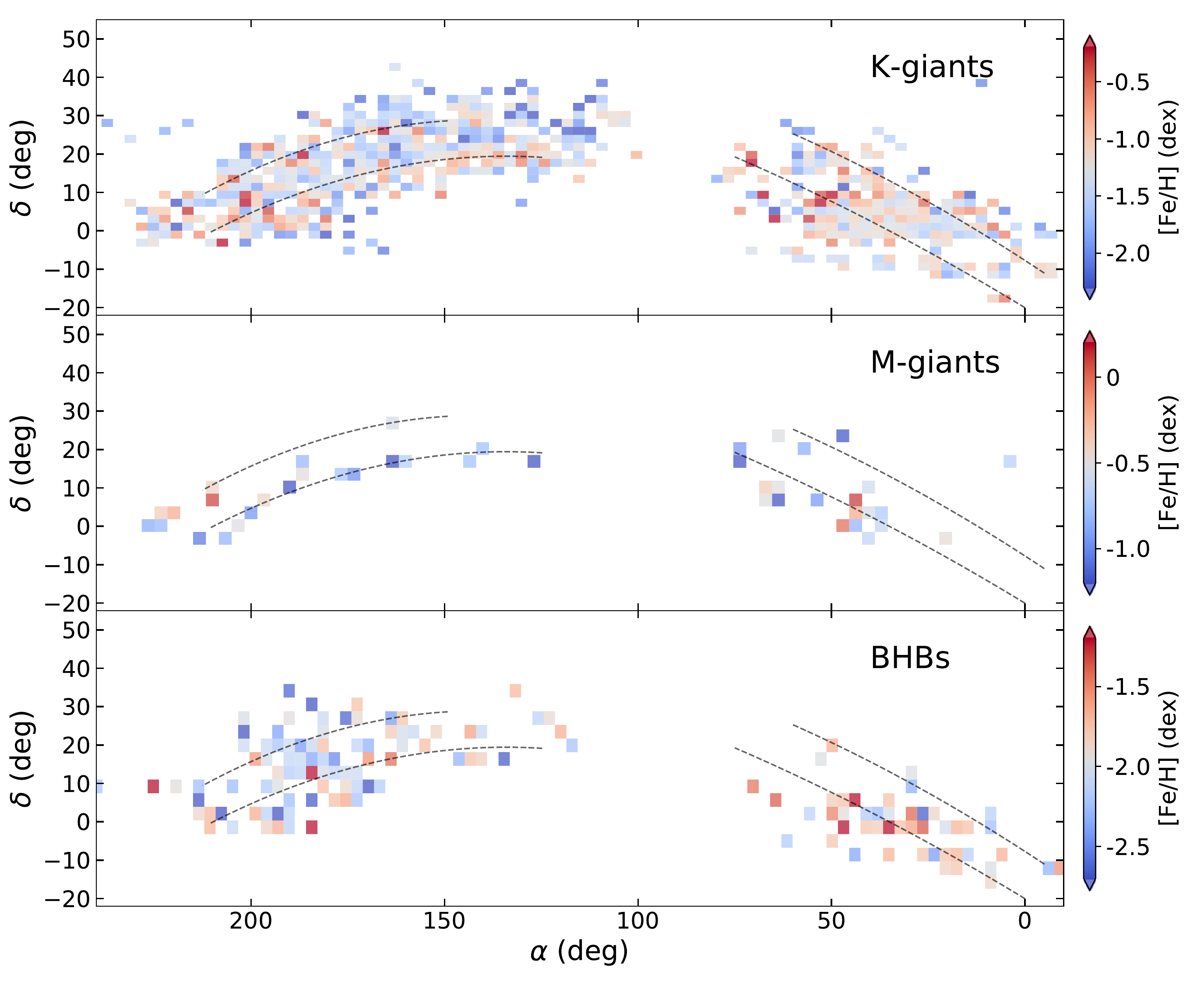}
  \end{tabular}
\caption{Density map of Sgr bifurcation from K-, M-giants and BHBs in equatorial coordinate color-coded according to metallicity. Dashed lines represent the center line of faint and bright stream.
} \label{bif_feh}
\end{figure}

\begin{figure}[htb]
\centering
  \begin{tabular}{@{}cccc@{}}
    \includegraphics[width=.5\textwidth]{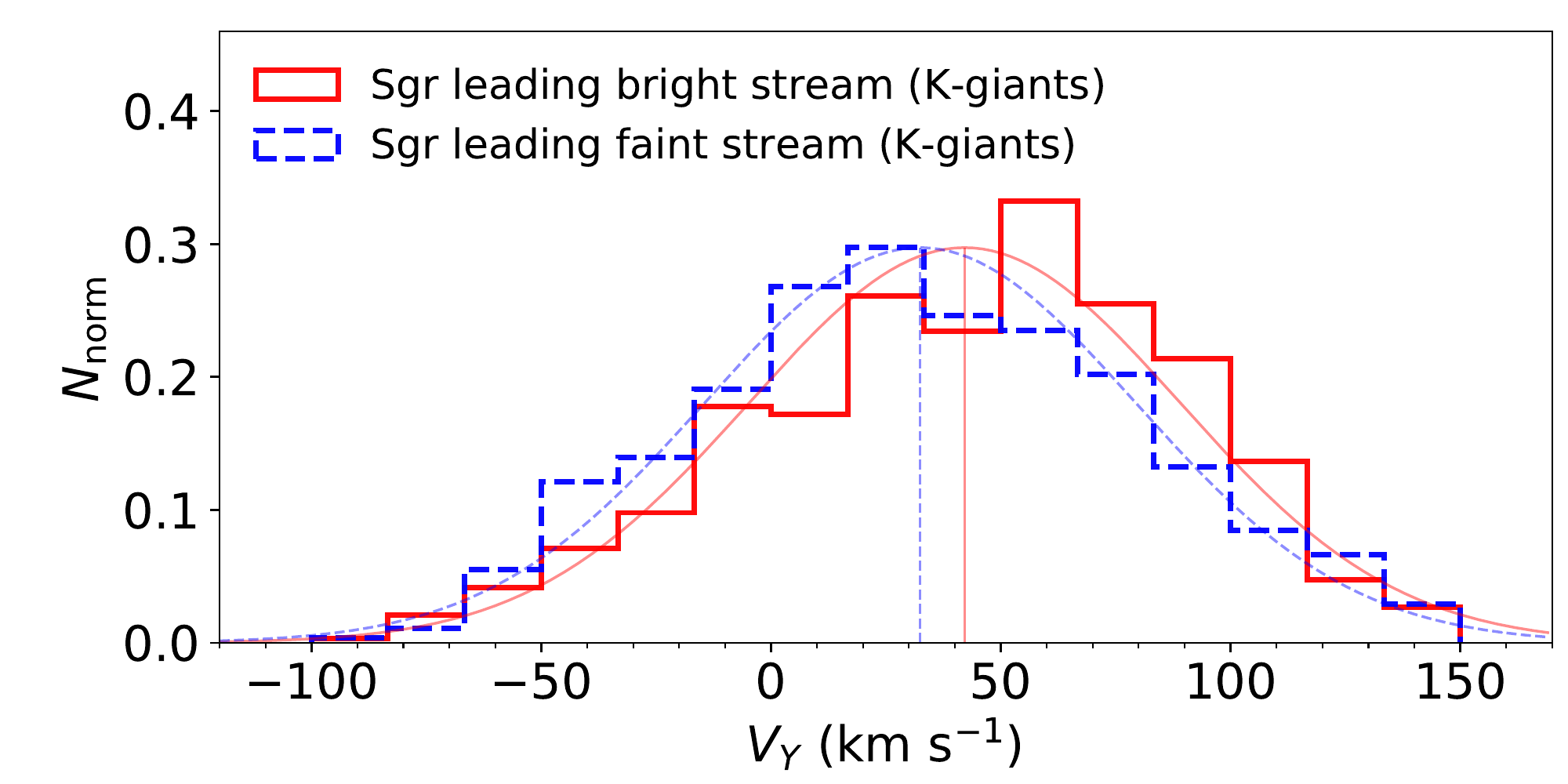} &
    \includegraphics[width=.5\textwidth]{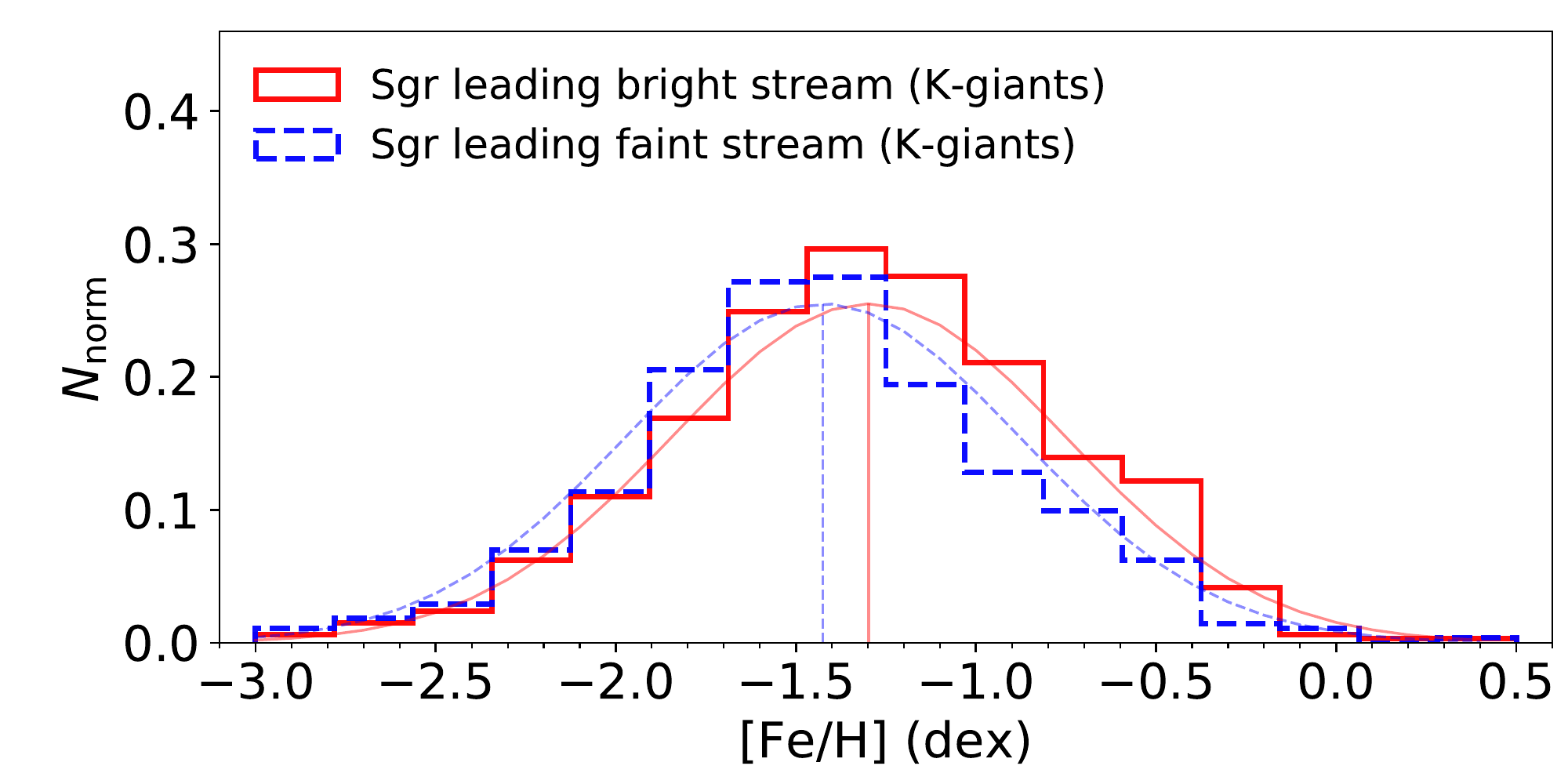} &\\
  \end{tabular}
  \begin{tabular}{@{}cccc@{}}
    \includegraphics[width=.5\textwidth]{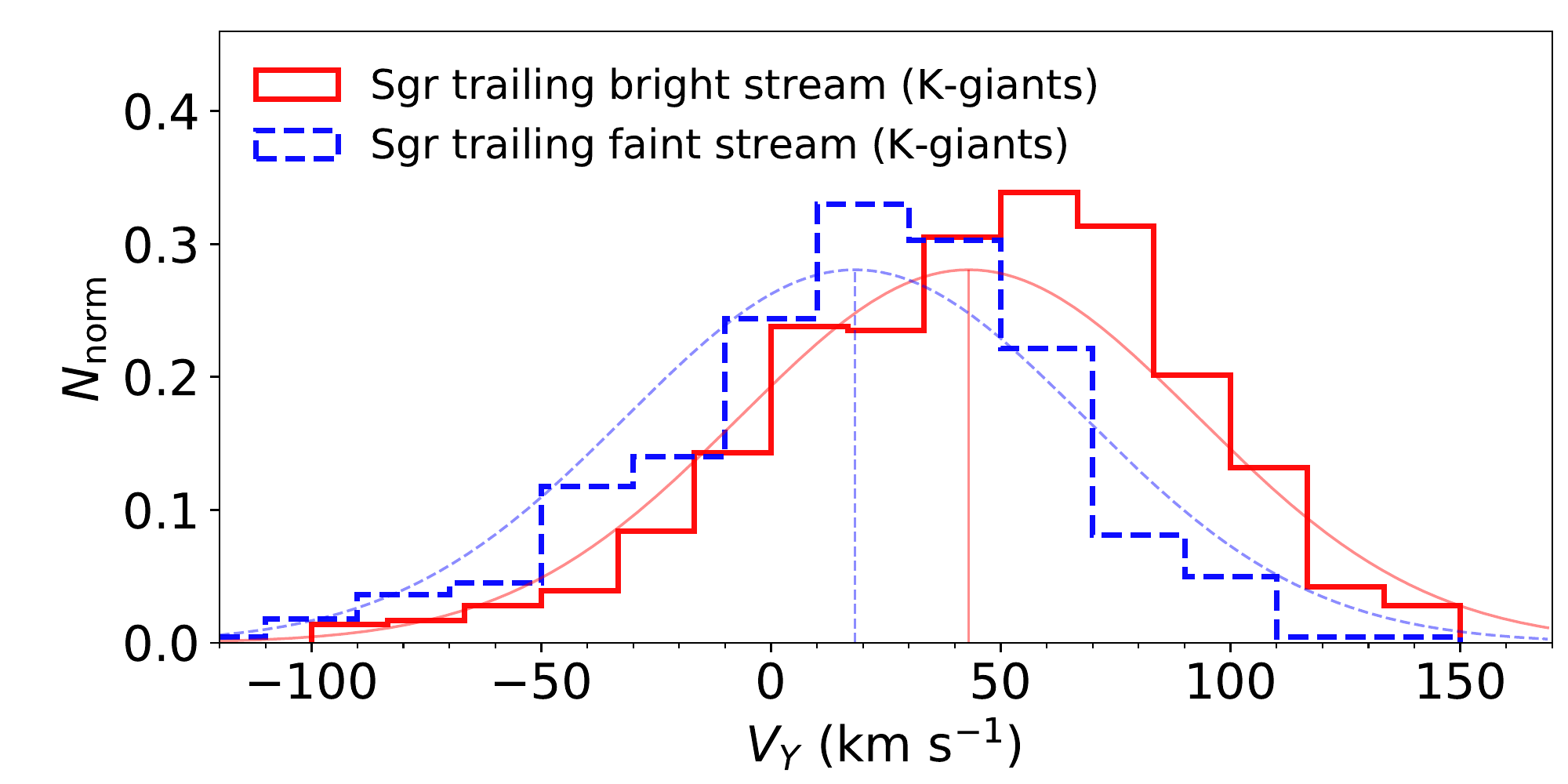} &
    \includegraphics[width=.5\textwidth]{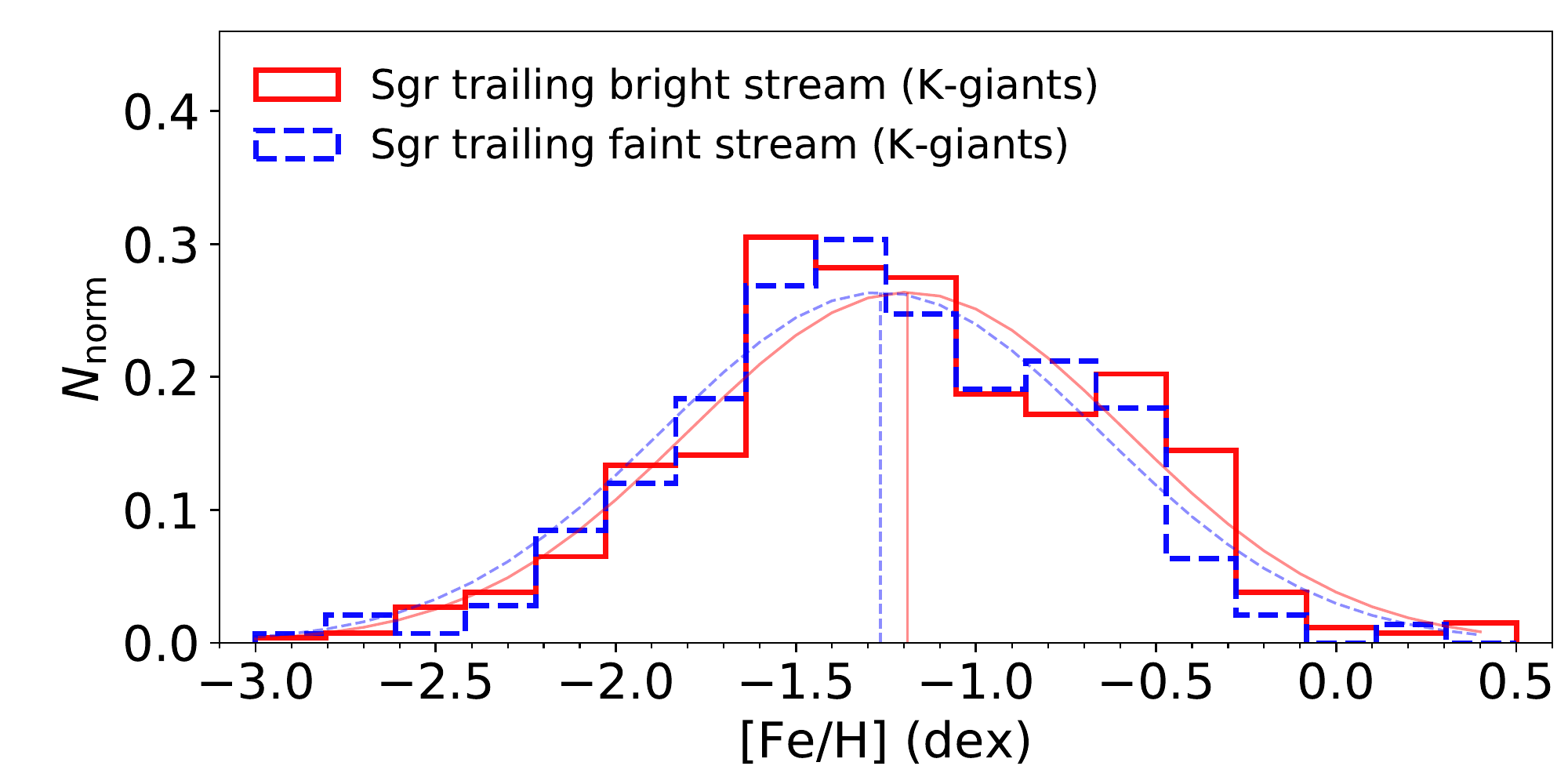} &\\
  \end{tabular}
\caption{Left and right panels respectively exhibit $V_Y$ and metallicity distribution of Sgr leading bright, faint stream and trailing bright, faint stream obtained by K-giants. Red histograms represent the bright stream. Blue histograms show the faint stream. Each histogram has a corresponding gaussian distribution obtained by mean value and scatter. In the left panels, $<V_Y>$ of leading bright and faint streams are 42.1 km s$^{-1}$ and 32.5 km s$^{-1}$, and those of trailing bright and faint streams are 43.0 km s$^{-1}$ and 18.3 km s$^{-1}$. In the right panels, $<\rm{[Fe/H]}>$ of leading bright and faint streams are $-$1.30 dex and $-$1.43 dex, and those of trailing bright and faint streams are $-$1.19 dex and $-$1.27 dex.
} \label{bif_kg}
\end{figure}

\begin{figure}[htb]
\centering
  \begin{tabular}{@{}cccc@{}}
    \includegraphics[width=.5\textwidth]{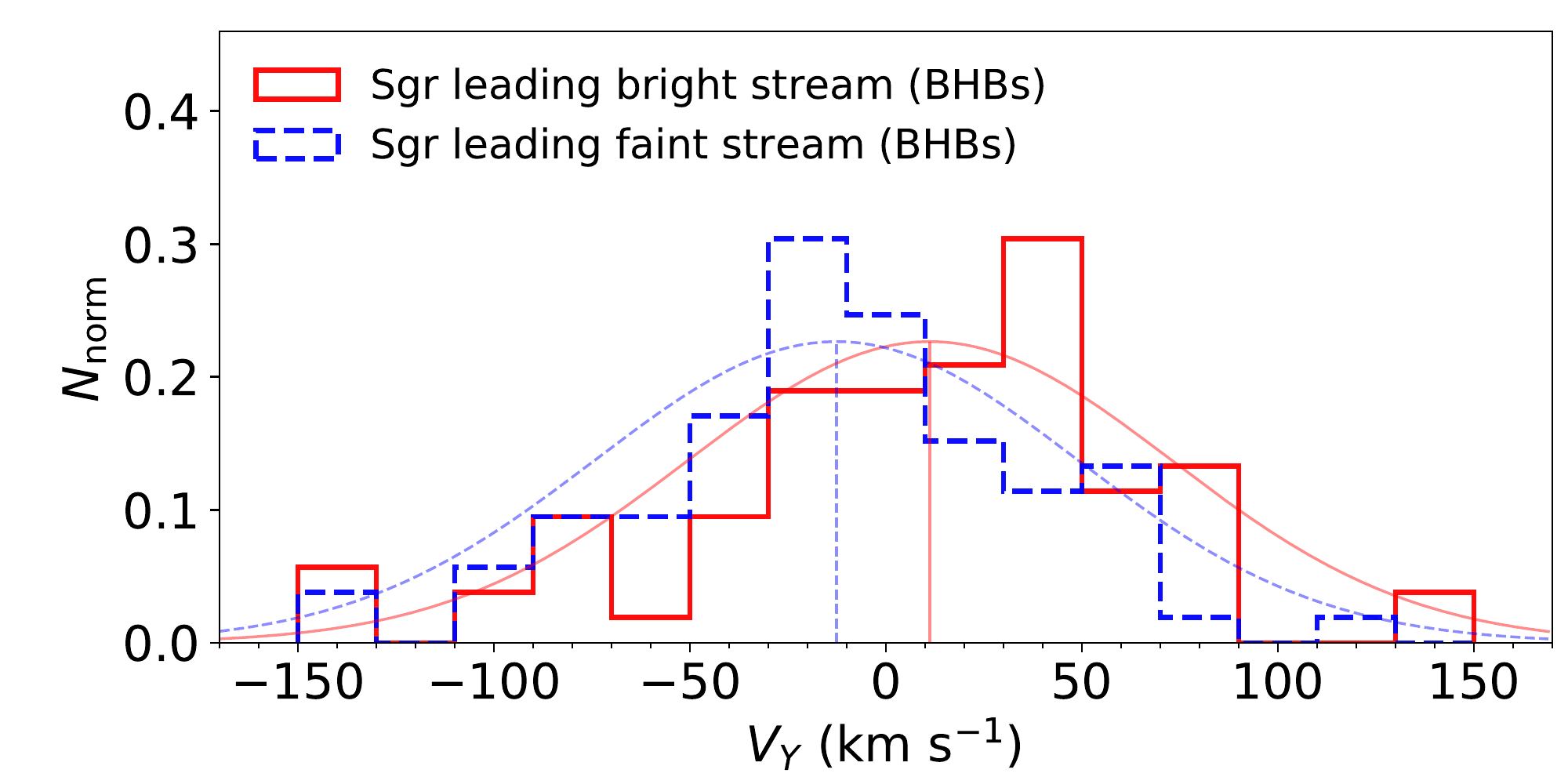} &
    \includegraphics[width=.5\textwidth]{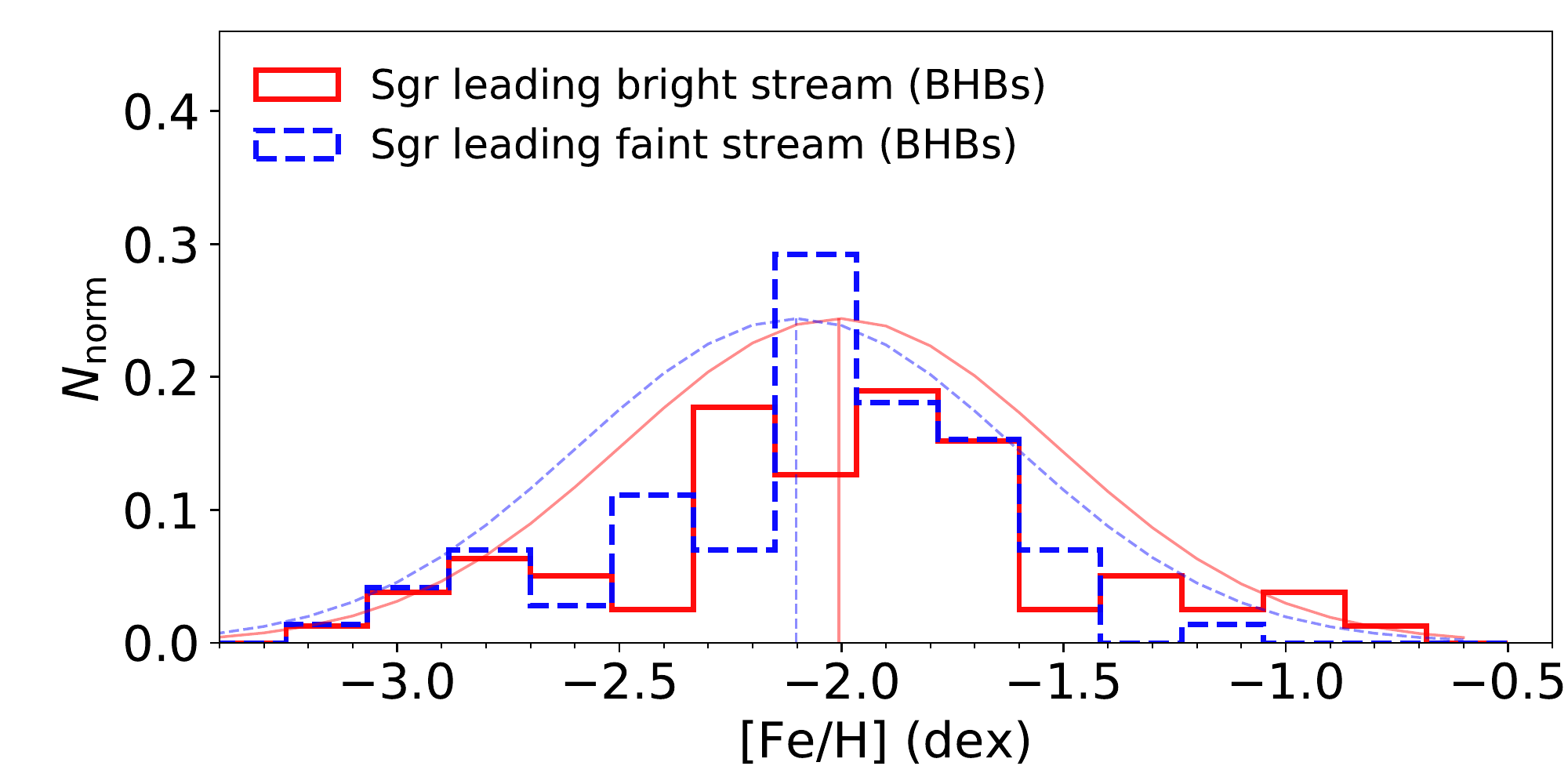} &\\
  \end{tabular}
  \begin{tabular}{@{}cccc@{}}
    \includegraphics[width=.5\textwidth]{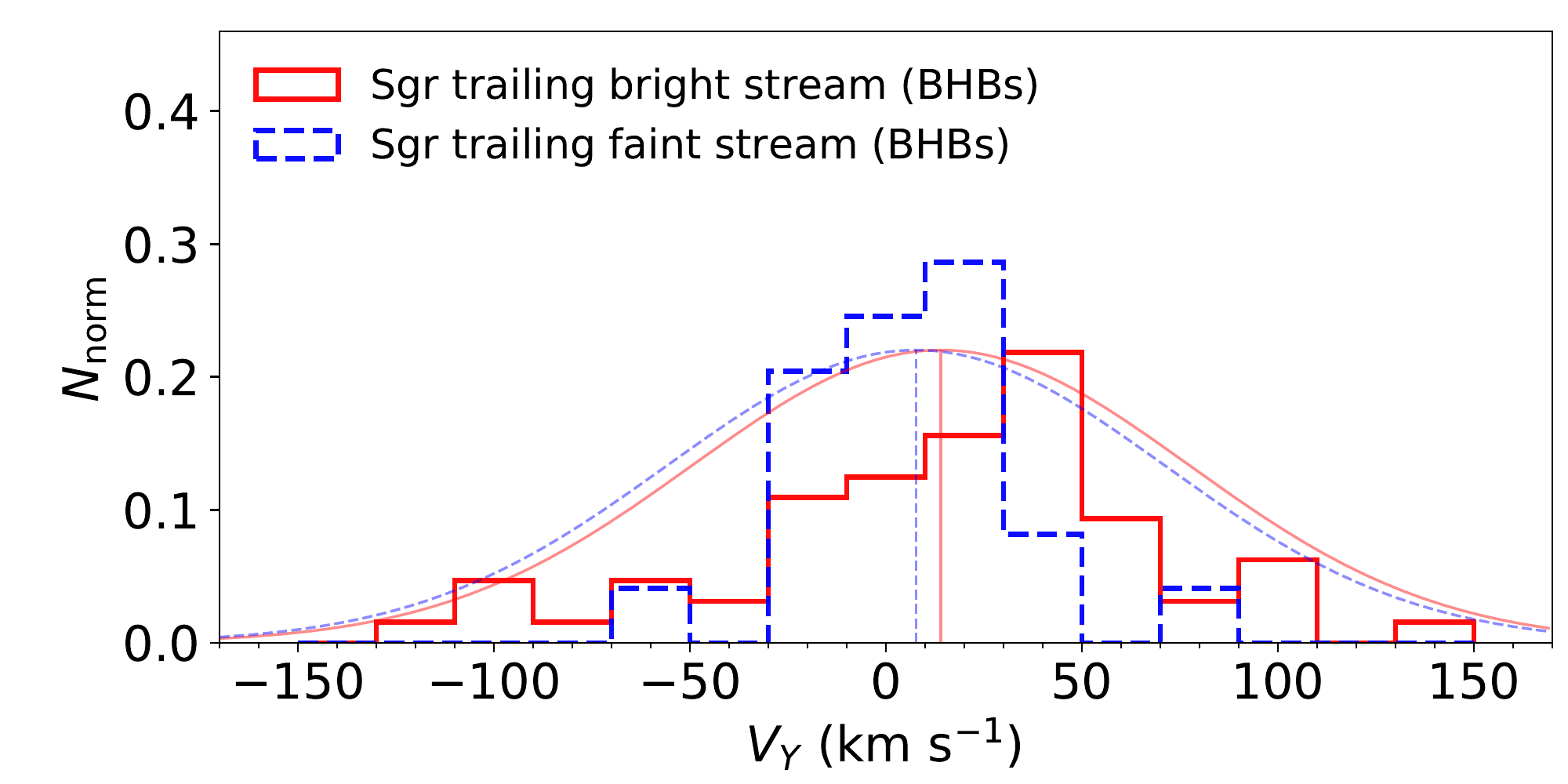} &
    \includegraphics[width=.5\textwidth]{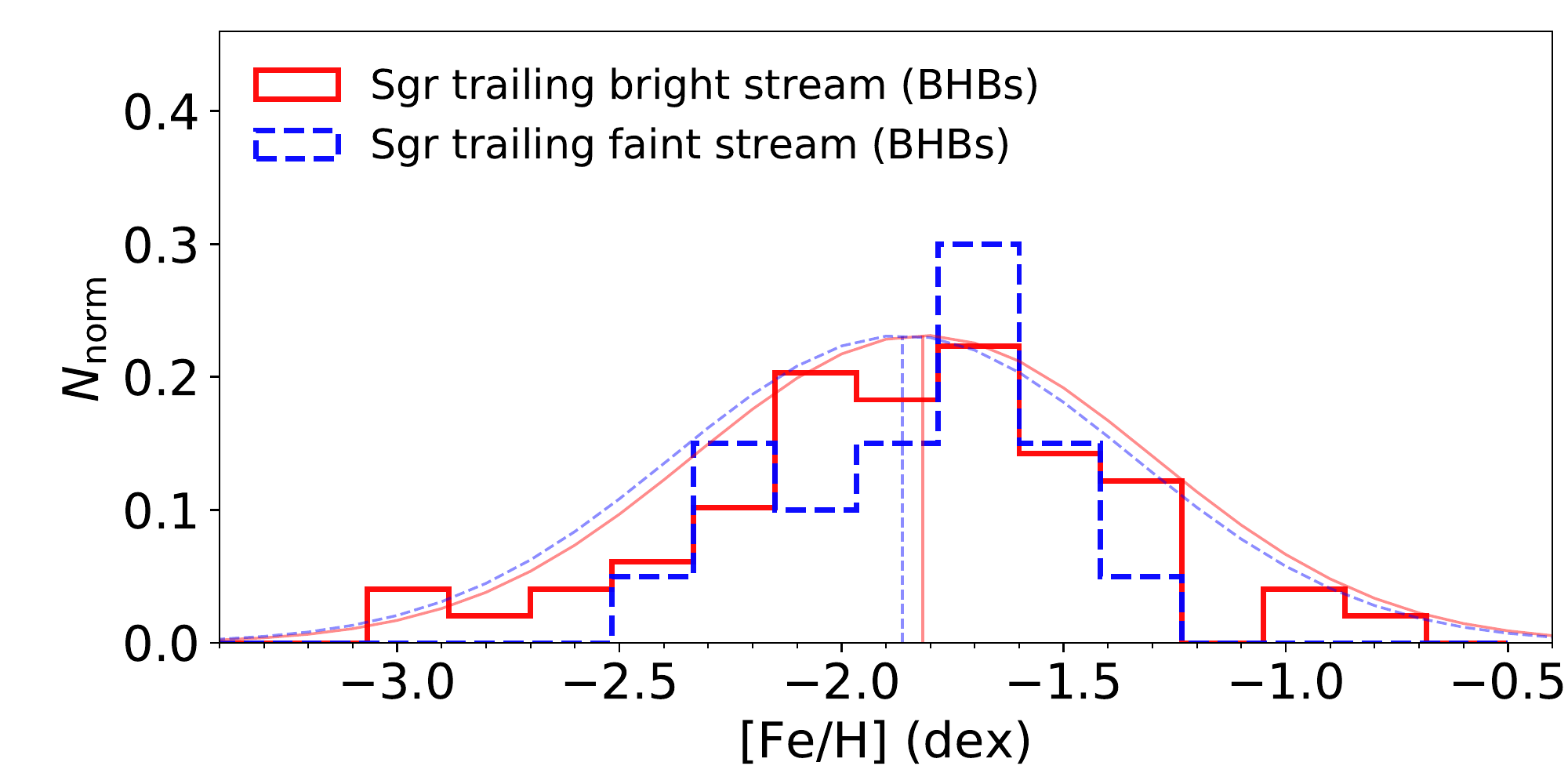} &\\
  \end{tabular}
\caption{$V_Y$ and metallicity distribution of Sgr bifurcation obtained by BHBs. In the left panels, the $<V_Y>$ of leading bright and faint streams are 11.2 km s$^{-1}$ and $-$12.6 km s$^{-1}$, and those of trailing bright and faint streams are 14.0 km s$^{-1}$ and 7.7 km s$^{-1}$. In the right panels, $<\rm{[Fe/H]}>$ of leading bright and faint streams are $-$2.01 dex and $-$2.10 dex, and those of trailing bright and faint streams are $-$1.82 dex and $-$1.86 dex.
} \label{bif_bhb}
\end{figure}

\begin{figure}[htb]
\centering
  \begin{tabular}{@{\hspace{-0.6cm}}c@{}}
  \includegraphics[width=0.7\textwidth]{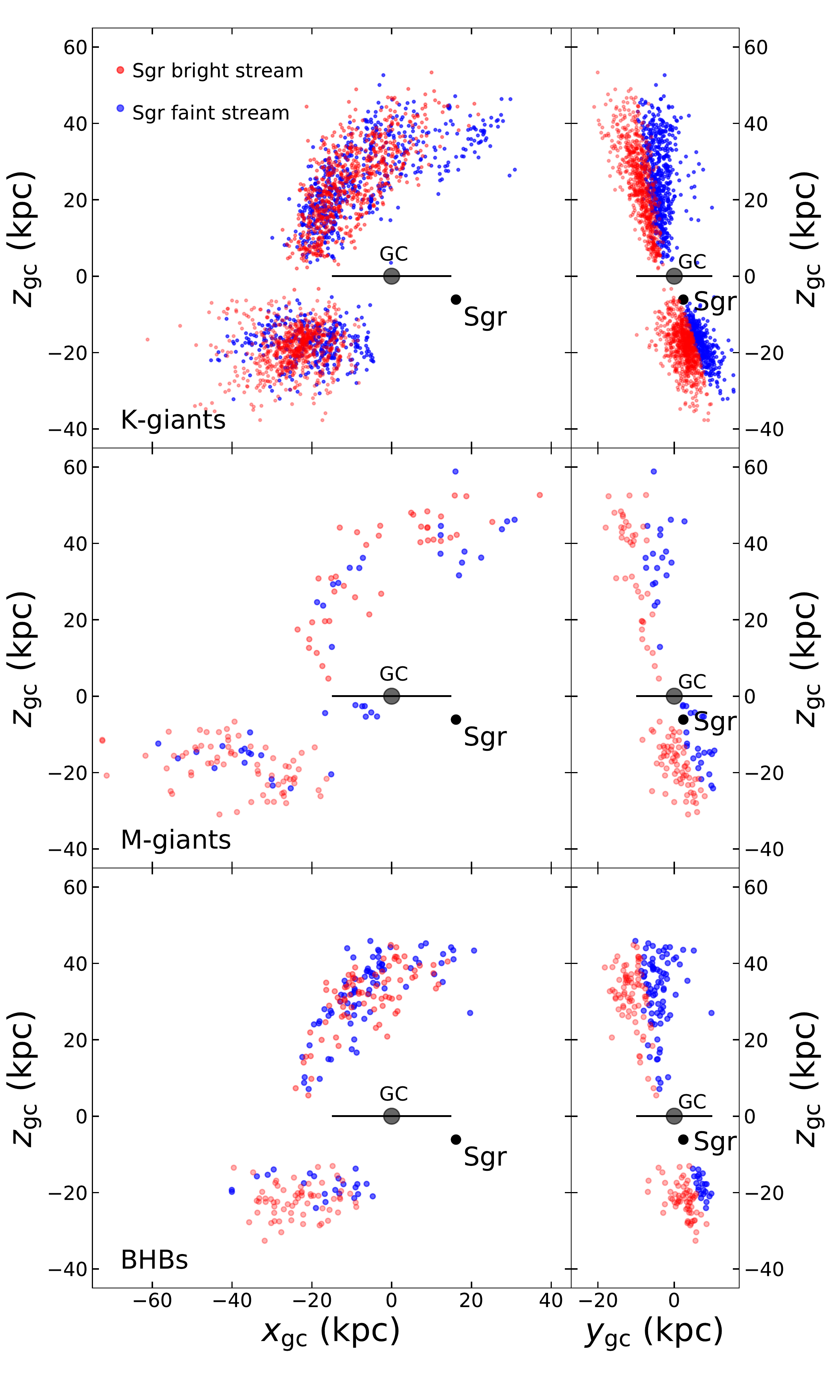}
  \end{tabular}
\caption{Sgr bifurcation from K-, M-giants and BHBs in $X$-$Z$ and $Y$-$Z$ plane, red and blue dots respectively indicate the bright stream stars and faint stream stars.
} \label{bif_xyz}
\end{figure}

In Section \ref{sub_sec:feh_vy}, we find the inner and outer side of Sgr stream are different in $V_Y$ and metallicity. Therefore, in Figure \ref{bif_vy} and \ref{bif_feh}, we present the density map of bifurcation with color-coded according to $V_Y$ and metallicity respectively. From the top panel of the Figure \ref{bif_vy} and \ref{bif_feh}, the K-giants show that faint and bright stream are also different in $V_Y$ and metallicity, bright stream is obviously higher in $V_Y$ and metallicity than faint stream, and in bottom panels of Figure \ref{bif_vy} and \ref{bif_feh}, BHBs also show a similar result. The M-giants members almost only cover on bright stream, which is consistent with the result found in \citet{Li16}. To examine the difference of $V_Y$ and metallicity between faint and bright stream appeared in K-giants and BHBs sample, in Figure \ref{bif_kg} and \ref{bif_bhb}, we divide the bifurcation into leading bright, faint stream and trailing bright, faint stream according to the dash-dotted line in Figure \ref{bif_div}. The result is both K-giants and BHBs present a same result, leading and trailing bright stream are on average higher in $V_Y$ and metallicity than those of leading and trailing faint stream. But the difference in metallicty is not as obvious as the velocity, especially trailing stream. In Figure \ref{bif_xyz}, we plot the divided bifurcation into $X$-$Z$ and $Y$-$Z$ plane. The figure shows that the faint and bright stream are two parallel stream along their moving direction. Thus, it is uncertainty that the $V_Y$ and metallicity difference between Sgr inner and outer side is related to Sgr bifurcation.

\section{Summary}\label{sec:summary}
By combining IoM and FoF algorithm, X19 picked up about 3,000 Sgr stream members from LAMOST, SDSS, and SEGUE-2, including K-giants, M-giants, and BHBs, which is the largest spectroscopic Sgr stream sample obtained yet. Based on this sample, we present the features of Sgr stream that we find.

We compare our Sgr sample with numerical simulations, \citetalias{LM10} and \citetalias{DL17}, and observation data from \citet{Koposov12}, \citet{Belokurov14} and \citet{Hernitschek17}. We find our sample is broadly consistent with \citetalias{DL17} model and observation data from \citet{Koposov12}, \citet{Belokurov14} and \citet{Hernitschek17}.

The velocity vector directions of Sgr debris and the low latitude M-giants in $X$-$Z$ plane indicate that the debris actually is the continuation of Sgr trailing stream and where the trailing stream stars return from the apocenter. Therefore, our sample shows that the apogalacticon of the Sgr trailing stream may reach $\sim$100 kpc from the Sun,  which is in agreement with previous observations like \citet{Belokurov14}, \citet{Koposov15}, \citet{Hernitschek17}, and \citet{Li19}. In addition, the energy versus angular momentum distribution of Sgr K-, M-giants, and BHBs shows no clear difference, but for Sgr streams, the debris and trailing stream are obviously higher in energy than leading stream.

We also present the metallicity distribution of Sgr K-, M-giants, and BHBs. M-giants is the most metal-rich population, followed by K-giants and BHBs. Additionally, the metallicities of Sgr leading, trailing, and debris are also different. All K-, M-giants and BHBs indicate that Sgr trailing stream is on average more metal-rich than leading stream, and K-giants show that Sgr debris is the most metal-poor population, which reflects their different unbound time from Sgr core. By comparing the $\alpha$-abundance of Sgr stars with the Galactic components and dwarf galaxies of the Milky Way, the trend of [$\alpha/$Fe] of Sgr stream is close to the Galactic halo at lower metallicity, then evolve down to lower [$\alpha/$Fe] than disk stars, and this evolution pattern is quite similar with Milky Way dwarf galaxies.

The $V_Y$ and metallicity distribution of Sgr stream in $X$-$Z$ plane shows that Sgr stream have a gradient along the line-of-sight direction from the Galactic center, the inner side of Sgr stream is higher in both $V_Y$ and metallicity, and $V_Y$ versus [Fe/H] shows that the $V_Y$ increases with metallicity, which means there indeed exists a correlation between $V_Y$ and metallicity. In addition, the Sgr bright and faint streams also exhibit different $V_Y$ and metallicity, with the bright stream higher in $V_Y$ and metallicity than the faint stream. But it is still hard to draw any conclusions that the $V_Y$ and metallicity difference between Sgr inner and outer side is related to Sgr bifurcation.

\acknowledgments
This study is supported by the National Natural Science Foundation of China under grants (NSFC) Nos. 11873052, 11890694, 11573032, and 11773033. J.L. acknowledges the NSFC under grant 11703019. L.Z. acknowledges support from NSFC grant 11703038. J.Z. would like to acknowledge the NSFC under grants U1731129. Q.F.X. thanks the NSFC for their support through grant 11603033. J.L.C. acknowledges support from HST grant HST-GO-15228 and NSF grant AST-1816196. This project was developed in part at the 2018 $Gaia$-LAMOST Sprint workshop supported by the NSFC under grants 11333003 and 11390372.

Guoshoujing Telescope (the Large Sky Area Multi-Object Fiber Spectroscopic Telescope LAMOST) is a National Major Scientific Project built by the Chinese Academy of Sciences. Funding for the project has been provided by the National Development and Reform Commission. LAMOST is operated and managed by the National Astronomical Observatories, Chinese Academy of Sciences.

This work has made use of data from the European Space Agency (ESA) mission {\it Gaia} (\url{https://www.cosmos.esa.int/gaia}), processed by the {\it Gaia} Data Processing and Analysis Consortium (DPAC, \url{https://www.cosmos.esa.int/web/gaia/dpac/consortium}). Funding for the DPAC has been provided by national institutions, in particular the institutions participating in the {\it Gaia} Multilateral Agreement.

\bibliographystyle{aasjournal}
\bibliography{Bibtex}

\begin{thebibliography}{}
\expandafter\ifx\csname natexlab\endcsname\relax\def\natexlab#1{#1}\fi
\providecommand{\url}[1]{\href{#1}{#1}}
\providecommand{\dodoi}[1]{doi:~\href{http://doi.org/#1}{\nolinkurl{#1}}}
\providecommand{\doeprint}[1]{\href{http://ascl.net/#1}{\nolinkurl{http://ascl.net/#1}}}
\providecommand{\doarXiv}[1]{\href{https://arxiv.org/abs/#1}{\nolinkurl{https://arxiv.org/abs/#1}}}

\bibitem[{{Bailer-Jones} {et~al.}(2018){Bailer-Jones}, {Rybizki}, {Fouesneau},
  {Mantelet}, \& {Andrae}}]{BJ18}
{Bailer-Jones}, C.~A.~L., {Rybizki}, J., {Fouesneau}, M., {Mantelet}, G., \&
  {Andrae}, R. 2018, \aj, 156, 58, \dodoi{10.3847/1538-3881/aacb21}

\bibitem[{{Belokurov} {et~al.}(2006){Belokurov}, {Zucker}, {Evans}, {Gilmore},
  {Vidrih}, {Bramich}, {Newberg}, {Wyse}, {Irwin}, {Fellhauer}, {Hewett},
  {Walton}, {Wilkinson}, {Cole}, {Yanny}, {Rockosi}, {Beers}, {Bell},
  {Brinkmann}, {Ivezi{\'c}}, \& {Lupton}}]{Belokurov06}
{Belokurov}, V., {Zucker}, D.~B., {Evans}, N.~W., {et~al.} 2006, \apjl, 642,
  L137, \dodoi{10.1086/504797}

\bibitem[{{Belokurov} {et~al.}(2014){Belokurov}, {Koposov}, {Evans},
  {Pe{\~n}arrubia}, {Irwin}, {Smith}, {Lewis}, {Gieles}, {Wilkinson},
  {Gilmore}, {Olszewski}, \& {Niederste-Ostholt}}]{Belokurov14}
{Belokurov}, V., {Koposov}, S.~E., {Evans}, N.~W., {et~al.} 2014, \mnras, 437,
  116, \dodoi{10.1093/mnras/stt1862}

\bibitem[{{Carlin} {et~al.}(2018){Carlin}, {Sheffield}, {Cunha}, \&
  {Smith}}]{Carlin18}
{Carlin}, J.~L., {Sheffield}, A.~A., {Cunha}, K., \& {Smith}, V.~V. 2018,
  \apjl, 859, L10, \dodoi{10.3847/2041-8213/aac3d8}

\bibitem[{{Carretta} {et~al.}(2010){Carretta}, {Bragaglia}, {Gratton},
  {Lucatello}, {Bellazzini}, {Catanzaro}, {Leone}, {Momany}, {Piotto}, \&
  {D'Orazi}}]{Carretta10}
{Carretta}, E., {Bragaglia}, A., {Gratton}, R.~G., {et~al.} 2010, \aap, 520,
  A95, \dodoi{10.1051/0004-6361/201014924}

\bibitem[{{Cohen} \& {Huang}(2009)}]{CH09}
{Cohen}, J.~G., \& {Huang}, W. 2009, \apj, 701, 1053,
  \dodoi{10.1088/0004-637X/701/2/1053}

\bibitem[{{Cui} {et~al.}(2012){Cui}, {Zhao}, {Chu}, {Li}, {Li}, {Zhang}, {Su},
  {Yao}, {Wang}, {Xing}, {Li}, {Zhu}, {Wang}, {Gu}, {Luo}, {Xu}, {Zhang},
  {Liu}, {Zhang}, {Yang}, {Cao}, {Chen}, {Chen}, {Chen}, {Chen}, {Chu}, {Feng},
  {Gong}, {Hou}, {Hu}, {Hu}, {Hu}, {Jia}, {Jiang}, {Jiang}, {Jiang}, {Jin},
  {Li}, {Li}, {Li}, {Liu}, {Liu}, {Lu}, {Mao}, {Men}, {Qi}, {Qi}, {Shi},
  {Tang}, {Tao}, {Wang}, {Wang}, {Wang}, {Wang}, {Wang}, {Wang}, {Wang},
  {Wang}, {Wang}, {Wang}, {Wang}, {Wang}, {Xu}, {Xu}, {Yang}, {Yu}, {Yuan},
  {Yuan}, {Zhai}, {Zhang}, {Zhang}, {Zhang}, {Zhao}, {Zhou}, {Zhou}, {Zhu}, \&
  {Zou}}]{Cui12}
{Cui}, X.-Q., {Zhao}, Y.-H., {Chu}, Y.-Q., {et~al.} 2012, Research in Astronomy
  and Astrophysics, 12, 1197, \dodoi{10.1088/1674-4527/12/9/003}

\bibitem[{{de Grijs} \& {Bono}(2016)}]{de_Grijs16}
{de Grijs}, R., \& {Bono}, G. 2016, \apjs, 227, 5,
  \dodoi{10.3847/0067-0049/227/1/5}

\bibitem[{{de Grijs} \& {Bono}(2017)}]{de_Grijs17}
---. 2017, \apjs, 232, 22, \dodoi{10.3847/1538-4365/aa8b71}

\bibitem[{{Dierickx} \& {Loeb}(2017)}]{DL17}
{Dierickx}, M.~I.~P., \& {Loeb}, A. 2017, \apj, 836, 92,
  \dodoi{10.3847/1538-4357/836/1/92}

\bibitem[{{Feuillet} {et~al.}(2016){Feuillet}, {Bovy}, {Holtzman}, {Girardi},
  {MacDonald}, {Majewski}, \& {Nidever}}]{Feuillet16}
{Feuillet}, D.~K., {Bovy}, J., {Holtzman}, J., {et~al.} 2016, \apj, 817, 40,
  \dodoi{10.3847/0004-637X/817/1/40}

\bibitem[{{Gaia Collaboration} {et~al.}(2018){Gaia Collaboration}, {Brown},
  {Vallenari}, {Prusti}, {de Bruijne}, {Babusiaux}, {Bailer-Jones}, {Biermann},
  {Evans}, {Eyer}, \& et~al.}]{Gaia18}
{Gaia Collaboration}, {Brown}, A.~G.~A., {Vallenari}, A., {et~al.} 2018, \aap,
  616, A1, \dodoi{10.1051/0004-6361/201833051}

\bibitem[{{Geisler} {et~al.}(2005){Geisler}, {Smith}, {Wallerstein},
  {Gonzalez}, \& {Charbonnel}}]{Geisler05}
{Geisler}, D., {Smith}, V.~V., {Wallerstein}, G., {Gonzalez}, G., \&
  {Charbonnel}, C. 2005, \aj, 129, 1428, \dodoi{10.1086/427540}

\bibitem[{{Hasselquist} {et~al.}(2019){Hasselquist}, {Carlin}, {Holtzman},
  {Shetrone}, {Hayes}, {Cunha}, {Smith}, {Beaton}, {Sobeck}, {Allende Prieto},
  {Majewski}, {Anguiano}, {Bizyaev}, {Garc{\'{\i}}a-Hern{\'a}ndez}, {Lane},
  {Pan}, {Nidever}, {Fern{\'a}ndez-Trincado}, {Wilson}, \&
  {Zamora}}]{Hasselquist19}
{Hasselquist}, S., {Carlin}, J.~L., {Holtzman}, J.~A., {et~al.} 2019, \apj,
  872, 58, \dodoi{10.3847/1538-4357/aafdac}

\bibitem[{{Hernitschek} {et~al.}(2017){Hernitschek}, {Sesar}, {Rix},
  {Belokurov}, {Martinez-Delgado}, {Martin}, {Kaiser}, {Hodapp}, {Chambers},
  {Wainscoat}, {Magnier}, {Kudritzki}, {Metcalfe}, \& {Draper}}]{Hernitschek17}
{Hernitschek}, N., {Sesar}, B., {Rix}, H.-W., {et~al.} 2017, \apj, 850, 96,
  \dodoi{10.3847/1538-4357/aa960c}

\bibitem[{{Hernquist}(1990)}]{Hernquist90}
{Hernquist}, L. 1990, \apj, 356, 359, \dodoi{10.1086/168845}

\bibitem[{{Holtzman} {et~al.}(2015){Holtzman}, {Shetrone}, {Johnson}, {Allende
  Prieto}, {Anders}, {Andrews}, {Beers}, {Bizyaev}, {Blanton}, {Bovy},
  {Carrera}, {Chojnowski}, {Cunha}, {Eisenstein}, {Feuillet}, {Frinchaboy},
  {Galbraith-Frew}, {Garc{\'{\i}}a P{\'e}rez}, {Garc{\'{\i}}a-Hern{\'a}ndez},
  {Hasselquist}, {Hayden}, {Hearty}, {Ivans}, {Majewski}, {Martell},
  {Meszaros}, {Muna}, {Nidever}, {Nguyen}, {O'Connell}, {Pan}, {Pinsonneault},
  {Robin}, {Schiavon}, {Shane}, {Sobeck}, {Smith}, {Troup}, {Weinberg},
  {Wilson}, {Wood-Vasey}, {Zamora}, \& {Zasowski}}]{Holtzman15}
{Holtzman}, J.~A., {Shetrone}, M., {Johnson}, J.~A., {et~al.} 2015, \aj, 150,
  148, \dodoi{10.1088/0004-6256/150/5/148}

\bibitem[{{Ibata} {et~al.}(2001){Ibata}, {Irwin}, {Lewis}, \&
  {Stolte}}]{Ibata01}
{Ibata}, R., {Irwin}, M., {Lewis}, G.~F., \& {Stolte}, A. 2001, \apjl, 547,
  L133, \dodoi{10.1086/318894}

\bibitem[{{Ibata} {et~al.}(1994){Ibata}, {Gilmore}, \& {Irwin}}]{Ibata94}
{Ibata}, R.~A., {Gilmore}, G., \& {Irwin}, M.~J. 1994, \nat, 370, 194,
  \dodoi{10.1038/370194a0}

\bibitem[{{Ibata} {et~al.}(1995){Ibata}, {Gilmore}, \& {Irwin}}]{Ibata95}
---. 1995, \mnras, 277, 781, \dodoi{10.1093/mnras/277.3.781}

\bibitem[{{Ibata} {et~al.}(1997){Ibata}, {Wyse}, {Gilmore}, {Irwin}, \&
  {Suntzeff}}]{Ibata97}
{Ibata}, R.~A., {Wyse}, R.~F.~G., {Gilmore}, G., {Irwin}, M.~J., \& {Suntzeff},
  N.~B. 1997, \aj, 113, 634, \dodoi{10.1086/118283}

\bibitem[{{Kirby} {et~al.}(2011){Kirby}, {Lanfranchi}, {Simon}, {Cohen}, \&
  {Guhathakurta}}]{Kirby11}
{Kirby}, E.~N., {Lanfranchi}, G.~A., {Simon}, J.~D., {Cohen}, J.~G., \&
  {Guhathakurta}, P. 2011, \apj, 727, 78, \dodoi{10.1088/0004-637X/727/2/78}

\bibitem[{{Koch} {et~al.}(2008){Koch}, {Grebel}, {Gilmore}, {Wyse}, {Kleyna},
  {Harbeck}, {Wilkinson}, \& {Evans}}]{Koch08}
{Koch}, A., {Grebel}, E.~K., {Gilmore}, G.~F., {et~al.} 2008, \aj, 135, 1580,
  \dodoi{10.1088/0004-6256/135/4/1580}

\bibitem[{{Koposov} {et~al.}(2015){Koposov}, {Belokurov}, {Zucker}, {Lewis},
  {Ibata}, {Olszewski}, {L{\'o}pez-S{\'a}nchez}, \& {Hyde}}]{Koposov15}
{Koposov}, S.~E., {Belokurov}, V., {Zucker}, D.~B., {et~al.} 2015, \mnras, 446,
  3110, \dodoi{10.1093/mnras/stu2263}

\bibitem[{{Koposov} {et~al.}(2012){Koposov}, {Belokurov}, {Evans}, {Gilmore},
  {Gieles}, {Irwin}, {Lewis}, {Niederste-Ostholt}, {Pe{\~n}arrubia}, {Smith},
  {Bizyaev}, {Malanushenko}, {Malanushenko}, {Schneider}, \&
  {Wyse}}]{Koposov12}
{Koposov}, S.~E., {Belokurov}, V., {Evans}, N.~W., {et~al.} 2012, \apj, 750,
  80, \dodoi{10.1088/0004-637X/750/1/80}

\bibitem[{{Law} \& {Majewski}(2010)}]{LM10}
{Law}, D.~R., \& {Majewski}, S.~R. 2010, \apj, 714, 229,
  \dodoi{10.1088/0004-637X/714/1/229}

\bibitem[{{Lee} {et~al.}(2008{\natexlab{a}}){Lee}, {Beers}, {Sivarani},
  {Allende Prieto}, {Koesterke}, {Wilhelm}, {Re Fiorentin}, {Bailer-Jones},
  {Norris}, {Rockosi}, {Yanny}, {Newberg}, {Covey}, {Zhang}, \& {Luo}}]{Lee08a}
{Lee}, Y.~S., {Beers}, T.~C., {Sivarani}, T., {et~al.} 2008{\natexlab{a}}, \aj,
  136, 2022, \dodoi{10.1088/0004-6256/136/5/2022}

\bibitem[{{Lee} {et~al.}(2008{\natexlab{b}}){Lee}, {Beers}, {Sivarani},
  {Johnson}, {An}, {Wilhelm}, {Allende Prieto}, {Koesterke}, {Re Fiorentin},
  {Bailer-Jones}, {Norris}, {Yanny}, {Rockosi}, {Newberg}, {Cudworth}, \&
  {Pan}}]{Lee08b}
---. 2008{\natexlab{b}}, \aj, 136, 2050, \dodoi{10.1088/0004-6256/136/5/2050}

\bibitem[{{Lemasle} {et~al.}(2012){Lemasle}, {Hill}, {Tolstoy}, {Venn},
  {Shetrone}, {Irwin}, {de Boer}, {Starkenburg}, \& {Salvadori}}]{Lemasle12}
{Lemasle}, B., {Hill}, V., {Tolstoy}, E., {et~al.} 2012, \aap, 538, A100,
  \dodoi{10.1051/0004-6361/201118132}

\bibitem[{{Lemasle} {et~al.}(2014){Lemasle}, {de Boer}, {Hill}, {Tolstoy},
  {Irwin}, {Jablonka}, {Venn}, {Battaglia}, {Starkenburg}, {Shetrone},
  {Letarte}, {Fran{\c c}ois}, {Helmi}, {Primas}, {Kaufer}, \&
  {Szeifert}}]{Lemasle14}
{Lemasle}, B., {de Boer}, T.~J.~L., {Hill}, V., {et~al.} 2014, \aap, 572, A88,
  \dodoi{10.1051/0004-6361/201423919}

\bibitem[{{Letarte} {et~al.}(2010){Letarte}, {Hill}, {Tolstoy}, {Jablonka},
  {Shetrone}, {Venn}, {Spite}, {Irwin}, {Battaglia}, {Helmi}, {Primas},
  {Fran{\c c}ois}, {Kaufer}, {Szeifert}, {Arimoto}, \& {Sadakane}}]{Letarte10}
{Letarte}, B., {Hill}, V., {Tolstoy}, E., {et~al.} 2010, \aap, 523, A17,
  \dodoi{10.1051/0004-6361/200913413}

\bibitem[{{Li} {et~al.}(2016){Li}, {Smith}, {Zhong}, {Hou}, {Carlin},
  {Newberg}, {Liu}, {Chen}, {Li}, {Shao}, {Small}, \& {Tian}}]{Li16}
{Li}, J., {Smith}, M.~C., {Zhong}, J., {et~al.} 2016, \apj, 823, 59,
  \dodoi{10.3847/0004-637X/823/1/59}

\bibitem[{{Li} {et~al.}(2019){Li}, {FELLOW}, {Liu}, {Xue}, {Zhong}, {Weiss},
  {Carlin}, {Tian}, \& {FELLOW}}]{Li19}
{Li}, J., {FELLOW}, ., {Liu}, C., {et~al.} 2019, \apj, 874, 138,
  \dodoi{10.3847/1538-4357/ab09ef}

\bibitem[{{Luo} {et~al.}(2012){Luo}, {Zhang}, {Zhao}, {Zhao}, {Cui}, {Li},
  {Chu}, {Shi}, {Wang}, {Zhang}, {Bai}, {Chen}, {Wang}, {Guo}, {Chen}, {Du},
  {Kong}, {Lei}, {Li}, {Song}, {Wu}, {Zhang}, {Zhou}, {Zuo}, {Du}, {He}, {Hou},
  {Dong}, {Li}, {Li}, {Li}, {Song}, {Tian}, {Wang}, {Wu}, {Yang}, {Yuan},
  {Cao}, {Chen}, {Chen}, {Chen}, {Chu}, {Feng}, {Gong}, {Gu}, {Hou}, {Huo},
  {Hu}, {Hu}, {Hu}, {Jia}, {Jiang}, {Jiang}, {Jiang}, {Jin}, {Li}, {Li}, {Li},
  {Li}, {Li}, {Liu}, {Liu}, {Liu}, {Lu}, {Lu}, {Luo}, {Mao}, {Men}, {Ni}, {Qi},
  {Qi}, {Shi}, {Su}, {Sun}, {Su}, {Tang}, {Tao}, {Tu}, {Wang}, {Wang}, {Wang},
  {Wang}, {Wang}, {Wang}, {Wang}, {Wang}, {Wang}, {Wang}, {Wang}, {Wang},
  {Wang}, {Wang}, {Wei}, {Xue}, {Xing}, {Xu}, {Xu}, {Xu}, {Yang}, {Yang},
  {Yao}, {Yu}, {Yuan}, {Zhai}, {Zhang}, {Zhang}, {Zhang}, {Zhang}, {Zhang},
  {Zhang}, {Zhao}, {Zhou}, {Zhu}, {Zhu}, \& {Zou}}]{Luo12}
{Luo}, A.-L., {Zhang}, H.-T., {Zhao}, Y.-H., {et~al.} 2012, Research in
  Astronomy and Astrophysics, 12, 1243, \dodoi{10.1088/1674-4527/12/9/004}

\bibitem[{{Majewski} {et~al.}(2004){Majewski}, {Ostheimer}, {Rocha-Pinto},
  {Patterson}, {Guhathakurta}, \& {Reitzel}}]{Majewski04}
{Majewski}, S.~R., {Ostheimer}, J.~C., {Rocha-Pinto}, H.~J., {et~al.} 2004,
  \apj, 615, 738, \dodoi{10.1086/424586}

\bibitem[{{Majewski} {et~al.}(1999){Majewski}, {Siegel}, {Kunkel}, {Reid},
  {Johnston}, {Thompson}, {Landolt}, \& {Palma}}]{Majewski99}
{Majewski}, S.~R., {Siegel}, M.~H., {Kunkel}, W.~E., {et~al.} 1999, \aj, 118,
  1709, \dodoi{10.1086/301036}

\bibitem[{{Majewski} {et~al.}(2003){Majewski}, {Skrutskie}, {Weinberg}, \&
  {Ostheimer}}]{Majewski03}
{Majewski}, S.~R., {Skrutskie}, M.~F., {Weinberg}, M.~D., \& {Ostheimer}, J.~C.
  2003, \apj, 599, 1082, \dodoi{10.1086/379504}

\bibitem[{{Majewski} {et~al.}(2017){Majewski}, {Schiavon}, {Frinchaboy},
  {Allende Prieto}, {Barkhouser}, {Bizyaev}, {Blank}, {Brunner}, {Burton},
  {Carrera}, {Chojnowski}, {Cunha}, {Epstein}, {Fitzgerald}, {Garc{\'{\i}}a
  P{\'e}rez}, {Hearty}, {Henderson}, {Holtzman}, {Johnson}, {Lam}, {Lawler},
  {Maseman}, {M{\'e}sz{\'a}ros}, {Nelson}, {Nguyen}, {Nidever}, {Pinsonneault},
  {Shetrone}, {Smee}, {Smith}, {Stolberg}, {Skrutskie}, {Walker}, {Wilson},
  {Zasowski}, {Anders}, {Basu}, {Beland}, {Blanton}, {Bovy}, {Brownstein},
  {Carlberg}, {Chaplin}, {Chiappini}, {Eisenstein}, {Elsworth}, {Feuillet},
  {Fleming}, {Galbraith-Frew}, {Garc{\'{\i}}a}, {Garc{\'{\i}}a-Hern{\'a}ndez},
  {Gillespie}, {Girardi}, {Gunn}, {Hasselquist}, {Hayden}, {Hekker}, {Ivans},
  {Kinemuchi}, {Klaene}, {Mahadevan}, {Mathur}, {Mosser}, {Muna}, {Munn},
  {Nichol}, {O'Connell}, {Parejko}, {Robin}, {Rocha-Pinto}, {Schultheis},
  {Serenelli}, {Shane}, {Silva Aguirre}, {Sobeck}, {Thompson}, {Troup},
  {Weinberg}, \& {Zamora}}]{Majewski17}
{Majewski}, S.~R., {Schiavon}, R.~P., {Frinchaboy}, P.~M., {et~al.} 2017, \aj,
  154, 94, \dodoi{10.3847/1538-3881/aa784d}

\bibitem[{{McWilliam} {et~al.}(2013){McWilliam}, {Wallerstein}, \&
  {Mottini}}]{McWilliam13}
{McWilliam}, A., {Wallerstein}, G., \& {Mottini}, M. 2013, \apj, 778, 149,
  \dodoi{10.1088/0004-637X/778/2/149}

\bibitem[{{Monaco} {et~al.}(2005){Monaco}, {Bellazzini}, {Bonifacio},
  {Ferraro}, {Marconi}, {Pancino}, {Sbordone}, \& {Zaggia}}]{Monaco05}
{Monaco}, L., {Bellazzini}, M., {Bonifacio}, P., {et~al.} 2005, \aap, 441, 141,
  \dodoi{10.1051/0004-6361:20053333}

\bibitem[{{Navarro} {et~al.}(1996){Navarro}, {Frenk}, \& {White}}]{NFW96}
{Navarro}, J.~F., {Frenk}, C.~S., \& {White}, S.~D.~M. 1996, \apj, 462, 563,
  \dodoi{10.1086/177173}

\bibitem[{{Newberg} \& {Carlin}(2016)}]{NC16}
{Newberg}, H.~J., \& {Carlin}, J.~L., eds. 2016, Astrophysics and Space Science
  Library, Vol. 420, {Tidal Streams in the Local Group and Beyond}

\bibitem[{{Newberg} {et~al.}(2002){Newberg}, {Yanny}, {Rockosi}, {Grebel},
  {Rix}, {Brinkmann}, {Csabai}, {Hennessy}, {Hindsley}, {Ibata}, {Ivezi{\'c}},
  {Lamb}, {Nash}, {Odenkirchen}, {Rave}, {Schneider}, {Smith}, {Stolte}, \&
  {York}}]{Newberg02}
{Newberg}, H.~J., {Yanny}, B., {Rockosi}, C., {et~al.} 2002, \apj, 569, 245,
  \dodoi{10.1086/338983}

\bibitem[{{Newberg} {et~al.}(2003){Newberg}, {Yanny}, {Grebel}, {Hennessy},
  {Ivezi{\'c}}, {Martinez-Delgado}, {Odenkirchen}, {Rix}, {Brinkmann}, {Lamb},
  {Schneider}, \& {York}}]{Newberg03}
{Newberg}, H.~J., {Yanny}, B., {Grebel}, E.~K., {et~al.} 2003, \apjl, 596,
  L191, \dodoi{10.1086/379316}

\bibitem[{{Nidever} {et~al.}(2019){Nidever}, {Hasselquist}, {Hayes}, {Hawkins},
  {Majewski}, {Smith}, {Anguiano}, {Povick}, {Stringfellow}, {Sobeck}, {Cunha},
  {Nitschelm}, {Fernandez-Trincado}, {Shetrone}, {Beers}, {Cohen}, {Allende
  Prieto}, {Gallart}, {Garcia-Hernandez}, {Dell'Agli}, {Jonsson}, \&
  {Lacerna}}]{Nidever19}
{Nidever}, D.~L., {Hasselquist}, S., {Hayes}, C.~R., {et~al.} 2019, arXiv
  e-prints.
\newblock \doarXiv{1901.03448}

\bibitem[{{Sbordone} {et~al.}(2007){Sbordone}, {Bonifacio}, {Buonanno},
  {Marconi}, {Monaco}, \& {Zaggia}}]{Sbordone07}
{Sbordone}, L., {Bonifacio}, P., {Buonanno}, R., {et~al.} 2007, \aap, 465, 815,
  \dodoi{10.1051/0004-6361:20066385}

\bibitem[{{Sch{\"o}nrich} {et~al.}(2010){Sch{\"o}nrich}, {Binney}, \&
  {Dehnen}}]{Schonrich10}
{Sch{\"o}nrich}, R., {Binney}, J., \& {Dehnen}, W. 2010, \mnras, 403, 1829,
  \dodoi{10.1111/j.1365-2966.2010.16253.x}

\bibitem[{{Sesar} {et~al.}(2017){Sesar}, {Hernitschek}, {Dierickx}, {Fardal},
  \& {Rix}}]{Sesarl17}
{Sesar}, B., {Hernitschek}, N., {Dierickx}, M.~I.~P., {Fardal}, M.~A., \&
  {Rix}, H.-W. 2017, \apjl, 844, L4, \dodoi{10.3847/2041-8213/aa7c61}

\bibitem[{{Shetrone} {et~al.}(2003){Shetrone}, {Venn}, {Tolstoy}, {Primas},
  {Hill}, \& {Kaufer}}]{Shetrone03}
{Shetrone}, M., {Venn}, K.~A., {Tolstoy}, E., {et~al.} 2003, \aj, 125, 684,
  \dodoi{10.1086/345966}

\bibitem[{{Shetrone} {et~al.}(2001){Shetrone}, {C{\^o}t{\'e}}, \&
  {Sargent}}]{Shetrone01}
{Shetrone}, M.~D., {C{\^o}t{\'e}}, P., \& {Sargent}, W.~L.~W. 2001, \apj, 548,
  592, \dodoi{10.1086/319022}

\bibitem[{{Siegel} {et~al.}(2007){Siegel}, {Dotter}, {Majewski}, {Sarajedini},
  {Chaboyer}, {Nidever}, {Anderson}, {Mar{\'{\i}}n-Franch}, {Rosenberg},
  {Bedin}, {Aparicio}, {King}, {Piotto}, \& {Reid}}]{Siegel07}
{Siegel}, M.~H., {Dotter}, A., {Majewski}, S.~R., {et~al.} 2007, \apjl, 667,
  L57, \dodoi{10.1086/522003}

\bibitem[{{Skrutskie} {et~al.}(2006){Skrutskie}, {Cutri}, {Stiening},
  {Weinberg}, {Schneider}, {Carpenter}, {Beichman}, {Capps}, {Chester},
  {Elias}, {Huchra}, {Liebert}, {Lonsdale}, {Monet}, {Price}, {Seitzer},
  {Jarrett}, {Kirkpatrick}, {Gizis}, {Howard}, {Evans}, {Fowler}, {Fullmer},
  {Hurt}, {Light}, {Kopan}, {Marsh}, {McCallon}, {Tam}, {Van Dyk}, \&
  {Wheelock}}]{Skrutskie06}
{Skrutskie}, M.~F., {Cutri}, R.~M., {Stiening}, R., {et~al.} 2006, \aj, 131,
  1163, \dodoi{10.1086/498708}

\bibitem[{{Tolstoy} {et~al.}(2009){Tolstoy}, {Hill}, \& {Tosi}}]{Tolstoy09}
{Tolstoy}, E., {Hill}, V., \& {Tosi}, M. 2009, \araa, 47, 371,
  \dodoi{10.1146/annurev-astro-082708-101650}

\bibitem[{{Venn} {et~al.}(2012){Venn}, {Shetrone}, {Irwin}, {Hill}, {Jablonka},
  {Tolstoy}, {Lemasle}, {Divell}, {Starkenburg}, {Letarte}, {Baldner},
  {Battaglia}, {Helmi}, {Kaufer}, \& {Primas}}]{Venn12}
{Venn}, K.~A., {Shetrone}, M.~D., {Irwin}, M.~J., {et~al.} 2012, \apj, 751,
  102, \dodoi{10.1088/0004-637X/751/2/102}

\bibitem[{{Wu} {et~al.}(2011){Wu}, {Luo}, {Li}, {Shi}, {Prugniel}, {Liang},
  {Zhao}, {Zhang}, {Bai}, {Wei}, {Dong}, {Zhang}, \& {Chen}}]{Wu11}
{Wu}, Y., {Luo}, A.-L., {Li}, H.-N., {et~al.} 2011, Research in Astronomy and
  Astrophysics, 11, 924, \dodoi{10.1088/1674-4527/11/8/006}

\bibitem[{{Xue} {et~al.}(2008){Xue}, {Rix}, {Zhao}, {Re Fiorentin}, {Naab},
  {Steinmetz}, {van den Bosch}, {Beers}, {Lee}, {Bell}, {Rockosi}, {Yanny},
  {Newberg}, {Wilhelm}, {Kang}, {Smith}, \& {Schneider}}]{Xue08}
{Xue}, X.~X., {Rix}, H.~W., {Zhao}, G., {et~al.} 2008, \apj, 684, 1143,
  \dodoi{10.1086/589500}

\bibitem[{{Xue} {et~al.}(2011){Xue}, {Rix}, {Yanny}, {Beers}, {Bell}, {Zhao},
  {Bullock}, {Johnston}, {Morrison}, {Rockosi}, {Koposov}, {Kang}, {Liu},
  {Luo}, {Lee}, \& {Weaver}}]{Xue11}
{Xue}, X.-X., {Rix}, H.-W., {Yanny}, B., {et~al.} 2011, \apj, 738, 79,
  \dodoi{10.1088/0004-637X/738/1/79}

\bibitem[{{Xue} {et~al.}(2014){Xue}, {Ma}, {Rix}, {Morrison}, {Harding},
  {Beers}, {Ivans}, {Jacobson}, {Johnson}, {Lee}, {Lucatello}, {Rockosi},
  {Sobeck}, {Yanny}, {Zhao}, \& {Allende Prieto}}]{Xue14}
{Xue}, X.-X., {Ma}, Z., {Rix}, H.-W., {et~al.} 2014, \apj, 784, 170,
  \dodoi{10.1088/0004-637X/784/2/170}

\bibitem[{{Yanny} {et~al.}(2000){Yanny}, {Newberg}, {Kent},
  {Laurent-Muehleisen}, {Pier}, {Richards}, {Stoughton}, {Anderson}, {Annis},
  {Brinkmann}, {Chen}, {Csabai}, {Doi}, {Fukugita}, {Hennessy}, {Ivezi{\'c}},
  {Knapp}, {Lupton}, {Munn}, {Nash}, {Rockosi}, {Schneider}, {Smith}, \&
  {York}}]{Yanny00}
{Yanny}, B., {Newberg}, H.~J., {Kent}, S., {et~al.} 2000, \apj, 540, 825,
  \dodoi{10.1086/309386}

\bibitem[{{Yanny} {et~al.}(2009){Yanny}, {Rockosi}, {Newberg}, {Knapp},
  {Adelman-McCarthy}, {Alcorn}, {Allam}, {Allende Prieto}, {An}, {Anderson},
  {Anderson}, {Bailer-Jones}, {Bastian}, {Beers}, {Bell}, {Belokurov},
  {Bizyaev}, {Blythe}, {Bochanski}, {Boroski}, {Brinchmann}, {Brinkmann},
  {Brewington}, {Carey}, {Cudworth}, {Evans}, {Evans}, {Gates}, {G{\"a}nsicke},
  {Gillespie}, {Gilmore}, {Gomez-Moran}, {Grebel}, {Greenwell}, {Gunn},
  {Jordan}, {Jordan}, {Harding}, {Harris}, {Hendry}, {Holder}, {Ivans},
  {Ivezi{\v c}}, {Jester}, {Johnson}, {Kent}, {Kleinman}, {Kniazev},
  {Krzesinski}, {Kron}, {Kuropatkin}, {Lebedeva}, {Lee}, {Leger}, {L{\'e}pine},
  {Levine}, {Lin}, {Long}, {Loomis}, {Lupton}, {Malanushenko}, {Malanushenko},
  {Margon}, {Martinez-Delgado}, {McGehee}, {Monet}, {Morrison}, {Munn},
  {Neilsen}, {Nitta}, {Norris}, {Oravetz}, {Owen}, {Padmanabhan}, {Pan},
  {Peterson}, {Pier}, {Platson}, {Fiorentin}, {Richards}, {Rix}, {Schlegel},
  {Schneider}, {Schreiber}, {Schwope}, {Sibley}, {Simmons}, {Snedden}, {Smith},
  {Stark}, {Stauffer}, {Steinmetz}, {Stoughton}, {Subba Rao}, {Szalay},
  {Szkody}, {Thakar}, {Thirupathi}, {Tucker}, {Uomoto}, {Vanden Berk},
  {Vidrih}, {Wadadekar}, {Watters}, {Wilhelm}, {Wyse}, {Yarger}, \&
  {Zucker}}]{Yanny09}
{Yanny}, B., {Rockosi}, C., {Newberg}, H.~J., {et~al.} 2009, \aj, 137, 4377,
  \dodoi{10.1088/0004-6256/137/5/4377}

\bibitem[{{York} {et~al.}(2000){York}, {Adelman}, {Anderson}, {Anderson},
  {Annis}, {Bahcall}, {Bakken}, {Barkhouser}, {Bastian}, {Berman}, {Boroski},
  {Bracker}, {Briegel}, {Briggs}, {Brinkmann}, {Brunner}, {Burles}, {Carey},
  {Carr}, {Castander}, {Chen}, {Colestock}, {Connolly}, {Crocker}, {Csabai},
  {Czarapata}, {Davis}, {Doi}, {Dombeck}, {Eisenstein}, {Ellman}, {Elms},
  {Evans}, {Fan}, {Federwitz}, {Fiscelli}, {Friedman}, {Frieman}, {Fukugita},
  {Gillespie}, {Gunn}, {Gurbani}, {de Haas}, {Haldeman}, {Harris}, {Hayes},
  {Heckman}, {Hennessy}, {Hindsley}, {Holm}, {Holmgren}, {Huang}, {Hull},
  {Husby}, {Ichikawa}, {Ichikawa}, {Ivezi{\'c}}, {Kent}, {Kim}, {Kinney},
  {Klaene}, {Kleinman}, {Kleinman}, {Knapp}, {Korienek}, {Kron}, {Kunszt},
  {Lamb}, {Lee}, {Leger}, {Limmongkol}, {Lindenmeyer}, {Long}, {Loomis},
  {Loveday}, {Lucinio}, {Lupton}, {MacKinnon}, {Mannery}, {Mantsch}, {Margon},
  {McGehee}, {McKay}, {Meiksin}, {Merelli}, {Monet}, {Munn}, {Narayanan},
  {Nash}, {Neilsen}, {Neswold}, {Newberg}, {Nichol}, {Nicinski}, {Nonino},
  {Okada}, {Okamura}, {Ostriker}, {Owen}, {Pauls}, {Peoples}, {Peterson},
  {Petravick}, {Pier}, {Pope}, {Pordes}, {Prosapio}, {Rechenmacher}, {Quinn},
  {Richards}, {Richmond}, {Rivetta}, {Rockosi}, {Ruthmansdorfer}, {Sandford},
  {Schlegel}, {Schneider}, {Sekiguchi}, {Sergey}, {Shimasaku}, {Siegmund},
  {Smee}, {Smith}, {Snedden}, {Stone}, {Stoughton}, {Strauss}, {Stubbs},
  {SubbaRao}, {Szalay}, {Szapudi}, {Szokoly}, {Thakar}, {Tremonti}, {Tucker},
  {Uomoto}, {Vanden Berk}, {Vogeley}, {Waddell}, {Wang}, {Watanabe},
  {Weinberg}, {Yanny}, {Yasuda}, \& {SDSS Collaboration}}]{York00}
{York}, D.~G., {Adelman}, J., {Anderson}, Jr., J.~E., {et~al.} 2000, \aj, 120,
  1579, \dodoi{10.1086/301513}

\bibitem[{{Zhang} {et~al.}(2019){Zhang}, {Liu}, \& {Deng}}]{Zhang19}
{Zhang}, B., {Liu}, C., \& {Deng}, L.-C. 2019, arXiv e-prints.
\newblock \doarXiv{1908.08677}

\bibitem[{{Zhao} {et~al.}(2012){Zhao}, {Zhao}, {Chu}, {Jing}, \&
  {Deng}}]{Zhao12}
{Zhao}, G., {Zhao}, Y.-H., {Chu}, Y.-Q., {Jing}, Y.-P., \& {Deng}, L.-C. 2012,
  Research in Astronomy and Astrophysics, 12, 723,
  \dodoi{10.1088/1674-4527/12/7/002}

\bibitem[{{Zhong} {et~al.}(2019){Zhong}, {Li}, {Carlin}, {Chen}, {Mendez}, \&
  {Hou}}]{Zhong19}
{Zhong}, J., {Li}, J., {Carlin}, J.~L., {et~al.} 2019, \apjs,
  \dodoi{10.3847/1538-4365/ab3859}

\bibitem[{{Zhong} {et~al.}(2015){Zhong}, {L{\'e}pine}, {Li}, {Chen}, {Hou},
  {Yang}, {Li}, {Zhang}, \& {Hou}}]{Zhong15}
{Zhong}, J., {L{\'e}pine}, S., {Li}, J., {et~al.} 2015, Research in Astronomy
  and Astrophysics, 15, 1154, \dodoi{10.1088/1674-4527/15/8/005}

\end{thebibliography}
\clearpage

\newpage
\vspace*{\fill}
\vfill
\clearpage

\newpage
\vspace*{\fill}
\vfill
\clearpage

\newpage
\vspace*{\fill}
\vfill
\clearpage

\newpage
\vspace*{\fill}
\vfill
\clearpage

\newpage
\vspace*{\fill}
\vfill
\clearpage

\newpage
\vspace*{\fill}
\vfill
\clearpage

\newpage
\vspace*{\fill}
\vfill
\clearpage

\newpage
\vspace*{\fill}
\vfill
\clearpage

\newpage
\vspace*{\fill}
\vfill
\clearpage

\newpage
\vspace*{\fill}
\vfill
\clearpage

\newpage
\vspace*{\fill}
\vfill
\clearpage

\newpage
\vspace*{\fill}
\vfill
\clearpage

\newpage
\vspace*{\fill}
\vfill
\clearpage

\newpage
\vspace*{\fill}
\vfill
\clearpage

\begin{sidewaystable}[htb]
\caption{Parameters of Sgr Stream Stars} \label{t_catalog}
\footnotesize
\setlength\tabcolsep{4pt}
\begin{tabular}{@{\extracolsep{-0.1cm}}cclrr ccr@{\hspace{0.3cm}}r@{\hspace{0.5cm}}c ccccc cc}
\tablewidth{0pt}
\hline
\hline
\colhead{LAMOST/SDSS\tablenotemark{a}}         &
\colhead{Gaia\tablenotemark{b}}                &
\colhead{Type}                &
\colhead{R.A.}                &
\colhead{Decl.}               &
\colhead{$d$}                 &
\colhead{$\delta d$}              &
\colhead{$hrv$}               &
\colhead{$\delta hrv$}            &
\colhead{pmra}                &
\colhead{$\delta$pmra}             &
\colhead{pmdec}               &
\colhead{$\delta$pmdec}            &
\colhead{[Fe/H]}              &
\colhead{$\delta$[Fe/H]}           &
\colhead{[$\alpha$/Fe]}       &
\colhead{$\delta [\alpha$/Fe]}
\\
\colhead{} &
\colhead{} &
\colhead{} &
\colhead{deg} &
\colhead{deg} &
\colhead{kpc} &
\colhead{kpc} &
\colhead{km s$^{-1}$} &
\colhead{km s$^{-1}$} &
\colhead{mas yr$^{-1}$} &
\colhead{mas yr$^{-1}$} &
\colhead{mas yr$^{-1}$} &
\colhead{mas yr$^{-1}$} &
\colhead{dex} &
\colhead{dex} &
\colhead{dex} &
\colhead{dex}
\\
\hline
120814182 & 3967103412613001728 & LAMOST KG & 170.253884 & 15.573170 & 24.4 & 3.7 & $   1.4$ &  5.0 & $-1.605$ & 0.095 & $-1.413$ & 0.069 & $-1.18$ & 0.17 & $-0.02$ & 0.06  \\
121008028 & 1267711072598547072 & LAMOST KG & 221.696725 & 26.496090 & 16.6 & 1.4 & $-136.1$ & 10.6 & $ 0.429$ & 0.079 & $-1.754$ & 0.094 & $-1.81$ & 0.09 & $ 0.44$ & 0.04  \\
121905054 & 3938773327292638848 & LAMOST KG & 198.322328 & 18.246219 & 20.4 & 1.0 & $ -59.2$ &  6.9 & $-0.089$ & 0.062 & $-1.353$ & 0.049 & $-2.01$ & 0.09 & $ 0.31$ & 0.04  \\
121905244 & 3937264934779562624 & LAMOST KG & 198.274650 & 18.005971 & 20.0 & 1.3 & $ -60.4$ &  6.1 & $-0.108$ & 0.051 & $-1.337$ & 0.044 & $-1.62$ & 0.11 & $ 0.37$ & 0.05  \\
121905245 & 3937271497489649792 & LAMOST KG & 198.268435 & 18.148787 & 20.5 & 1.3 & $ -60.5$ &  7.8 & $-0.211$ & 0.078 & $-1.434$ & 0.060 & $-1.94$ & 0.15 & $ 0.39$ & 0.06  \\
123011177 & 1442795139541724544 & LAMOST KG & 200.612511 & 22.731429 & 21.1 & 0.8 & $ -44.8$ &  9.5 & $-0.188$ & 0.055 & $-1.384$ & 0.051 & $-2.11$ & 0.10 & $ 0.31$ & 0.04  \\
125604054 & 3882438096696454528 & LAMOST KG & 157.016266 & 10.747924 & 28.4 & 1.1 & $  22.8$ &  7.8 & $-1.207$ & 0.099 & $-0.999$ & 0.124 & $-1.62$ & 0.15 & $ 0.24$ & 0.06  \\
132101163 & 3918753557013284608 & LAMOST KG & 181.490768 & 11.441141 & 31.8 & 0.9 & $ -31.8$ &  8.6 & $-1.545$ & 0.061 & $-1.028$ & 0.036 & $-1.61$ & 0.08 & $ 0.40$ & 0.04  \\
132105204 & 3918794788699300992 & LAMOST KG & 181.092298 & 11.555653 & 43.1 & 2.0 & $ -29.7$ &  7.7 & $-1.041$ & 0.097 & $-0.785$ & 0.063 & $-1.59$ & 0.15 & $ 0.19$ & 0.06  \\
132109226 & 3920550124653418624 & LAMOST KG & 183.333534 & 13.041779 & 29.4 & 2.0 & $ -28.5$ &  6.2 & $-1.697$ & 0.074 & $-1.017$ & 0.055 & $-1.73$ & 0.12 & $ 0.19$ & 0.05  \\
\hline
\multicolumn{4}{l}{\textsuperscript{\hspace{-0.5em} a} Unique identifier in LAMOST/SDSS.}\\
\multicolumn{4}{l}{\textsuperscript{\hspace{-0.5em} b} Solution identifier in Gaia.}\\
\multicolumn{4}{l}{(This table is available in its entirety in machine-readable form.)}
\end{tabular}
\end{sidewaystable}

\begin{sidewaystable}[htb]
\caption{Orbital Parameters of Sgr Stream Stars} \label{t_orbs}
\scriptsize 
\setlength\tabcolsep{10pt}
\begin{tabular}{ccccc ccrrc ccccc}
\tablewidth{0pt}
\hline
\hline
\colhead{LAMOST/SDSS} &
\colhead{$e$} &
\colhead{$\delta e$} &
\colhead{$a$} &
\colhead{$\delta a$} &
\colhead{$l_{\rm{orb}}$} &
\colhead{$\delta l_{\rm{orb}}$} &
\colhead{$b_{\rm{orb}}$} &
\colhead{$\delta b_{\rm{orb}}$} &
\colhead{$l_{\rm{apo}}$} &
\colhead{$\delta l_{\rm{apo}}$} &
\colhead{$E$} &
\colhead{$\delta E$} &
\colhead{$L$}&
\colhead{$\delta L$}
\\
\colhead{ } &
\colhead{ } &
\colhead{ } &
\colhead{kpc} &
\colhead{kpc} &
\colhead{deg} &
\colhead{deg} &
\colhead{deg} &
\colhead{deg} &
\colhead{deg} &
\colhead{deg} &
\colhead{km$^2$ s$^{-2}$} &
\colhead{km$^2$ s$^{-2}$} &
\colhead{km s$^{-1}$ kpc} &
\colhead{km s$^{-1}$ kpc}
\\
\hline
120814182 & 0.46 & 0.10 & 19.79 & 5.02 & 277.27 & 13.35 & 106.41 & 7.23 & 115.22 & 12.80 & $-76520.25$ & 6485.40 & 3212.90 & 975.30 \\
121008028 & 0.47 & 0.03 & 15.34 & 1.68 & 159.99 &  6.49 & 104.02 & 1.19 & 311.62 &  8.98 & $-85054.91$ & 3722.07 & 2548.25 & 234.91 \\
121905054 & 0.55 & 0.03 & 16.45 & 0.77 & 157.44 &  3.91 & 100.70 & 1.51 & 304.42 &  4.87 & $-82276.59$ & 1607.59 & 2499.43 & 123.88 \\
121905244 & 0.55 & 0.03 & 16.21 & 0.84 & 159.80 &  4.27 & 101.28 & 1.91 & 306.31 &  4.62 & $-82777.67$ & 1830.66 & 2465.24 & 106.50 \\
121905245 & 0.61 & 0.04 & 15.63 & 0.88 & 157.13 &  5.66 & 100.45 & 2.03 & 309.11 &  6.35 & $-83764.70$ & 1969.44 & 2206.07 & 149.70 \\
123011177 & 0.51 & 0.03 & 15.63 & 0.58 & 151.47 &  3.85 & 101.95 & 0.99 & 308.07 &  5.24 & $-84242.28$ & 1270.22 & 2487.49 & 124.20 \\
125604054 & 0.38 & 0.07 & 26.05 & 2.14 & 268.87 &  7.24 & 121.74 & 4.24 & 100.11 & 12.84 & $-67939.22$ & 2322.00 & 4318.00 & 500.08 \\
132101163 & 0.37 & 0.04 & 28.84 & 2.13 & 283.61 &  3.34 & 104.90 & 1.05 &  90.38 &  5.61 & $-64795.94$ & 2170.06 & 4785.16 & 412.04 \\
132105204 & 0.49 & 0.09 & 35.77 & 4.40 & 283.98 &  8.01 & 105.67 & 2.11 &  94.79 & 10.74 & $-57819.74$ & 3114.20 & 5274.65 & 991.59 \\
132109226 & 0.31 & 0.06 & 26.13 & 3.52 & 278.76 &  6.32 & 104.13 & 1.84 &  88.22 &  7.14 & $-68086.86$ & 3893.76 & 4517.30 & 667.87 \\
\hline
\multicolumn{8}{l}{(This table is available in its entirety in machine-readable form.)}
\end{tabular}
\end{sidewaystable}

\vspace*{\fill}
\begin{table}[htb]
\centering
\caption{Coordinates of the Sgr Trailing Stream Bifurcation} \label{t_bif}
\footnotesize
\setlength\tabcolsep{15pt}
\begin{tabular}{ccc}
\hline
\hline
\colhead{R.A.} &
\colhead{Decl. (cen, bright)} &
\colhead{Decl. (cen, faint)}
\vspace{-0.5em}
\\
\colhead{(deg)} &
\colhead{(deg)} &
\colhead{(deg)}
\\
\hline
$-5.0$ &   ...   & $-10.0$ \\
$ 0.0$ & $-20.0$ & $ -8.0$ \\
$ 5.0$ & $-17.0$ & $ -5.0$ \\
$10.0$ & $-14.0$ & $ -2.0$ \\
$15.0$ & $-11.0$ & $  2.5$ \\
$20.0$ & $ -8.5$ & $  5.0$ \\
$25.0$ & $ -5.5$ & $  8.0$ \\
$30.0$ & $ -3.0$ & $ 10.0$ \\
$35.0$ & $  0.5$ & $ 12.5$ \\
$40.0$ & $  2.5$ & $ 15.5$ \\
$45.0$ & $  5.0$ & $ 18.5$ \\
$50.0$ & $  7.5$ & $ 20.5$ \\
$55.0$ & $ 10.5$ & $ 23.0$ \\
$60.0$ & $ 12.5$ & $ 25.5$ \\
$65.0$ & $ 14.5$ &   ...   \\
$70.0$ & $ 17.0$ &   ...   \\
$75.0$ & $ 19.5$ &   ...   \\
\hline
\end{tabular}
\end{table}
\vspace*{\fill}

\end{document}